\documentclass[journal]{IEEEtran}

\usepackage{stfloats}
\usepackage{algorithm}  
\usepackage{algpseudocode}  
\usepackage{amsmath}

\usepackage{graphicx}
\usepackage[]{algpseudocode}
\usepackage{algorithmicx,algorithm}
\usepackage{amsmath}
\usepackage{titlesec}
\usepackage{multirow}  
\usepackage{amsthm,amsmath,amssymb}
\usepackage{graphicx}
\usepackage{float}
\usepackage{subfigure}
\usepackage{mathrsfs}
\usepackage{url}
\usepackage{array}
\newcolumntype{C}[1]{>{\centering}p{#1}}
\usepackage{makecell}
\usepackage{diagbox}
\usepackage{amsmath}

\usepackage{gensymb}
\usepackage{cite}
\usepackage{color}

\ifCLASSOPTIONcompsoc
 \usepackage[caption=false,font=normalsize,labelfont=sf,textfont=sf]{subfig}
\else
 \usepackage[caption=false,font=footnotesize]{subfig}
\fi

\UseRawInputEncoding
\begin{document}

\newtheorem{theorem}{Theorem}[section] 
\newtheorem{definition}[theorem]{Definition} 
\newtheorem{lemma}{Lemma} 
\newtheorem{corollary}[theorem]{Corollary}
\newtheorem{example}{Example}[section]
\newtheorem{proposition}[theorem]{Proposition}

\vspace{-15mm}
\title{Electromagnetic Property Sensing: A New Paradigm of Integrated Sensing and Communication}


\author{   
Yuhua Jiang, Feifei Gao, and Shi Jin



\thanks{
Y. Jiang and F. Gao are with Institute for Artificial Intelligence, Tsinghua University (THUAI), 
State Key Lab of Intelligent Technologies and Systems, Tsinghua University, 
Beijing National Research Center for Information Science and Technology (BNRist), Beijing, P.R. China (email: jiangyh20@mails.tsinghua.edu.cn, feifeigao@ieee.org).


S. Jin is with the National Mobile Communications Research 
Laboratory, Southeast University, Nanjing 210096, China (e-mail: jinshi@seu.edu.cn).


}
}

\maketitle
\vspace{-15mm}
\begin{abstract}



Integrated sensing and communication (ISAC) has 
opened up numerous game-changing opportunities for future wireless systems.
In this paper, we develop a novel scheme that utilizes orthogonal frequency division multiplexing (OFDM) pilot signals in ISAC systems to sense the electromagnetic (EM) property of the target and thus also identify the material of the target.
Specifically, we first establish an end-to-end EM propagation model by means of Maxwell equations, where the EM property of the target is captured by a closed-form expression of the ISAC channel, incorporating the Lippmann-Schwinger equation and the method of moments (MOM) for discretization. 
We then model the relative permittivity and conductivity distribution (RPCD) within a specified detection region.
Based on the sensing model, we introduce a multi-frequency-based EM property sensing method by which the RPCD can be reconstructed from compressive sensing techniques that exploits the joint sparsity structure of the EM property vector.  
To improve the sensing accuracy, we design a beamforming strategy from the communications transmitter based on the Born approximation that can minimize the mutual coherence of the sensing matrix.
The optimization problem is cast in terms of the Gram matrix and is solved iteratively to obtain the optimal beamforming matrix. 
Simulation results demonstrate the efficacy of the proposed method in achieving high-quality RPCD reconstruction and accurate material classification. Furthermore, improvements in RPCD reconstruction quality and material classification accuracy are observed with increased signal-to-noise ratio (SNR) or reduced target-transmitter distance.  
\end{abstract}


\IEEEpeerreviewmaketitle


\section{Introduction} 
The electromagnetic (EM) property serves as a crucial indicator to identify the material of targets, playing a pivotal role in various industries and enabling automation processes to operate efficiently and accurately \cite{nature1}.
In conventional material identification techniques, 
the infrared emissivity as a material specific property is investigated to increase the material classification  reliability \cite{nature1}.
However, the conventional infrared material identification techniques 
typically demand specialized devices that are costly and challenging to manufacture and maintain \cite{infrared1}. 
Besides, infrared light faces challenges in penetrating the target, particularly when the target surface is coated with camouflage material, rendering conventional infrared sensing methods ineffective. 
In contrast, the capability of EM waves to penetrate objects makes EM property sensing a viable and promising solution for accurate material identification. 
Besides, EM property sensing technology can be used in inspection and imaging of a human body, which meets the increasing demands from social security, environmental monitoring, and medical applications.


Recently, integrated sensing and communication (ISAC) has opened up numerous game-changing opportunities for future wireless sensing systems. 
ISAC allows communications systems and 
sensing systems to share the scarce radio spectrum, which saves a large amount of resource cost \cite{isac8,isac9,isac7,gaoisac}. 
In contrast to the dedicated sensing or communication functionality, the ISAC design methodology exhibits two types of gains. 
Firstly, the shared use of limited resources, namely, spectrum, energy, and hardware platforms, offers improved efficiency for both sensing and communications (S\&C), and thus provides integration gain. 
Secondly, mutual assistance between S\&C may offer coordination gain and further boost the dual performance.
Hence, ISAC is expected to benefit both communication and sensing functionality in the near future \cite{isac1,mypaper3,isac2,isac3}.
Due to its numerous advantages, ISAC is envisioned to be a key enabler for many future applications including intelligent connected vehicles, Internet of Things (IoT), and smart homes and cities \cite{isac4,isac5,isac6,isac10}.

While ISAC has achieved notable success in localization \cite{isac8}, tracking \cite{isac3}, imaging \cite{RIS_image}, and various other applications \cite{isac1,isac2,mypaper,isac4}, the realm of EM property sensing within ISAC remains unexplored to the best knowledge of the authors. 
Since the EM property of the target material is implicitly encoded into the channel state information (CSI) from the transmitter to the receiver, there are promising prospects to incorporate EM property sensing capabilities into the ISAC paradigm.
\textcolor{black}{
In the realm of ISAC, the primary function of sensing is to create an accurate representation that bridges the real physical world with its digital twin counterpart \cite{twin}, \cite{twin2}.
In this context, the sensing of EM properties becomes essential.
Unlike digital twins in image-based applications that may primarily rely on shape and location, digital twins in communications would need to reconstruct the communications channel, necessitating detailed material information.
Thus, accurate EM property data is crucial to construct digital twin scenarios precisely, as it provides the material characteristics essential for channel reconstruction.
This ensures that the digital representation not only mirrors the physical world's form but also its interactive characteristics. 
}




\textcolor{black}{
In this paper, we develop a novel scheme that utilizes OFDM pilot signals in ISAC systems to sense the EM property and identify the material of the target. }
The main contributions are summarized as follows:
\begin{itemize}
\item[$\bullet$] ISAC EM property sensing formulation: We establish an end-to-end EM propagation model from the perspective of Maxwell equations.    
A closed-form expression for a vector related to material EM property is derived, incorporating the Lippmann-Schwinger equation and the method of moments for discretization.
The objective is to reconstruct the relative permittivity and conductivity distribution (RPCD) within a specified detection region that contains the target.
\item[$\bullet$] Compressive sensing-based RPCD reconstruction: We propose a multi-frequency-based EM property sensing and material identification method using compressive sensing techniques, where the joint sparsity of the EM property vector is exploited to reconstruct RPCD.
\item[$\bullet$] Beamforming design for enhanced sensing: To further improve the sensing accuracy, we 
optimize the sensing matrix by designing the transmitter beamforming matrix based on the Born approximation. The design involves minimizing the difference between the Gram matrix of the designed beamforming matrix and a target matrix. 
The optimization problem is cast in terms of the Gram matrix and is solved iteratively to obtain the optimal beamforming matrix. 
\end{itemize}
Simulation results demonstrate that the proposed method can achieve both high-quality RPCD reconstruction and accurate material classification.
Moreover, the RPCD reconstruction quality and material classification accuracy can be improved by increasing the signal-to-noise ratio (SNR) at the receiver, or by decreasing the distance between the target and the transmitter.




The rest of this paper is organized as follows. 
Section~\uppercase\expandafter{\romannumeral2} presents the system model and formulates the EM property sensing problem.
Section~\uppercase\expandafter{\romannumeral3} derives the end-to-end electromagnetic propagation formula.
Section~\uppercase\expandafter{\romannumeral4} elaborates the strategy of sensing target material with multiple measurements.
Section~\uppercase\expandafter{\romannumeral5} describes the transmitter beamforming design criterion.
Section~\uppercase\expandafter{\romannumeral6} provides the extensive numerical simulation results, and 
Section~\uppercase\expandafter{\romannumeral7} draws the conclusion.

Notations: Boldface denotes vector or matrix;  $j$ corresponds to the imaginary unit; $(\cdot)^H$, $(\cdot)^T$, and $(\cdot)^*$ represent the Hermitian, transpose, and conjugate, respectively; 
$\circ$ denotes the Khatri-Rao product; 
$\odot$ denotes the Hadamard product;
$\otimes$ denotes the Kronecker product;
$\mathrm{vec}(\cdot)$ denotes the vectorization operation; 
 $\mathbf{A}[:,i]$ denotes the submatrix
composed of all elements in column $i$ of matrix $\mathbf{A}$;
$\mathbf{I}$, $\mathbf{1}$, and $\mathbf{0}$
denote the identity matrix, the all-ones vector, and the all-zeros vector with compatible dimensions;   
$\left\Vert\mathbf{a}\right\Vert_2$ denotes $\ell2$-norm of the vector $\mathbf{a}$; 
$\left|a\right|$ denotes the absolute value of the complex number $a$;
$\Re(\cdot)$ and $\Im(\cdot)$ denote the real and imaginary part of complex vectors or matrices, respectively; 
$\left\Vert\mathbf{A}\right\Vert_F$ denotes Frobenius-norm of the matrix $\mathbf{A}$; 
$\mathbf{A} \succeq \textbf{0}$ indicates that the matrix $\mathbf{A}$ is positive semi-definite. 
The distribution of a circularly symmetric complex Gaussian (CSCG) random vector with zero mean and 
covariance matrix $\mathbf{A}$ is denoted as $\mathcal{C N} (\mathbf{0}, \mathbf{A})$.





 \section{System Model and Problem Formulation}

\begin{figure}[t]
  \centering  \centerline{\includegraphics[width=8.9cm]{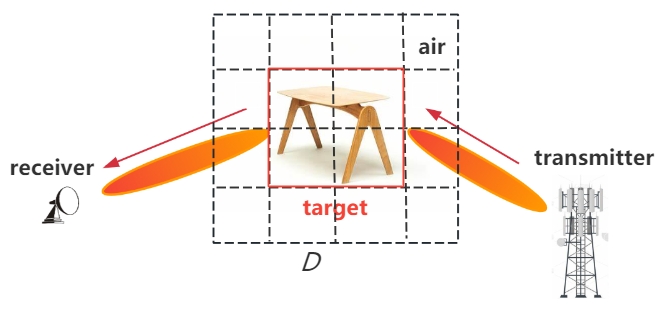}}
  \caption{The system model is composed of a multi-antenna transmitter, a target to be sensed, and a multi-antenna receiver. 
  }
  \label{system_model}
\end{figure}



Consider a downlink communications scenario, in which an $N_t$-antenna base station (BS) communicates with 
an $N_r$-antenna mobile user 
with orthogonal frequency division multiplexing (OFDM) signaling. 
The transmitter adopts fully digital precoding structure
where the number of RF chains $N_{RF}$ is equal to the
number of antennas $N_t$.  
To enable EM property sensing, we resort to the pilot transmission, in which a total of $K$ subcarriers are used to transmit pilots. 
The central frequency and the subcarrier interval of OFDM signals are $f_c$ and $\Delta f$, respectively. 
Then the transmission bandwidth is $ W = (K-1)\Delta f$, and the
frequency of the $k$-th subcarrier is $f_k = f_c + (2k-K-1)\Delta f/2$, where $k = 1, 2,\cdots, K$.
The corresponding wavelength of the $k$-th subcarrier is 
$\lambda_k = \frac{c}{f_c + (2k-K-1)\Delta f/2}$.
Furthermore, we consider that a total of $I$ pilot symbols are transmitted in each subcarrier.


\textcolor{black}{ 
The comprehensive sensing of a target in ISAC  is accomplished through the following steps.
Firstly, the existence of the target is ascertained through target detection. This initial phase is focused on discovering the presence of the target, which is a fundamental prerequisite for any further analysis.
Secondly, once the target is detected, the task of ISAC is pinpointing the precise location and velocity of the target.
These parameters are crucial as they provide the spatial context needed to accurately engage with the target. 
The methods to acquire these parameters have been comprehensively discussed in conventional ISAC literature  \cite{isac1,isac2,isac3,isac4,mypaper,isac5,isac6,isac7}.
Thirdly, we proceed to sense the EM property of the target, acquiring the material-specific information.    
Therefore, when sensing the EM property of the target, we may assume that other properties like the position of the target have already been obtained and are thus known values. 
}


\textcolor{black}{
Suppose the target scatters the pilot signals from the transmitter to the receiver as shown in Fig.~\ref{system_model}. 
Since only the signals scattered by the target carry the information of its EM property, we may send the pilot signals towards the target by beamforming properly at the transmitter. 
}  
The overall ISAC channel from the transmitter to the receiver of the $k$-th subcarrier hence only consists of the scattering path channel $\mathbf{H}_{k} \in \mathbb{C}^{N_r \times  N_t}$.
Thus, the received signals can be formulated as:
\begin{align}
\tilde{\mathbf{y}}_k = \mathbf{H}_{k} \tilde{\mathbf{w}}_k x_k  +  \tilde{\mathbf{n}}_k,
\end{align}
where $\tilde{\mathbf{w}}_k \in \mathbb{C}^{N_t \times  1}$ is the digital beamforming weights on all transmitting antennas; $x_k$ is the normalized pilot symbol; 
$\tilde{\mathbf{n}}_k \sim \mathcal{C N}\left(\mathbf{0}, \sigma_k^2 \mathbf{I}_{N_r}\right)$ is the complex Gaussian noise at the receiver of the $k$-th subcarrier.
For notational brevity, we denote $\mathbf{w}_k = \tilde{\mathbf{w}}_k x_k$ that is perfectly known by the receiver during the pilot transmission process.





Assume prior knowledge has determined that the target is contained within a certain sensing domain $D$, which can be discretized into a total of $M$ sampling points.
Let the vectors $\mathbf{E}_k^{i,D} \in \mathbb{C}^{M \times 1}$ and $\mathbf{J}_k^{D} \in \mathbb{C}^{M \times 1}$ collect the incident electric field and the equivalent contrast source at all $M$ sampling points in $D$, respectively \cite{lipp,m-born,v-born}.
Define $\mathbf{H}_{1,k} \in \mathbb{C}^{M \times N_t}$ 
as the channel matrix of the transmitter-to-target path that maps $\mathbf{w}_k$ to $\mathbf{E}_k^{i,D}$, and define 
$\mathbf{H}_{2,k} \in \mathbb{C}^{N_r \times M}$ as the channel matrix of the target-to-receiver path that maps $\mathbf{J}_k^{D}$
to $\tilde{\mathbf{y}}_k$.
The ISAC channel from the transmitter through the domain $D$
to the receiver can then be described as:
\begin{align}
\mathbf{H}_{k} = \mathbf{H}_{2,k}  
\mathbf{X}_k \mathbf{H}_{1,k}, 
\label{X}
\end{align}
where $\mathbf{X}_k \in \mathbb{C}^{M \times M}$ maps $\mathbf{E}_k^{i,D}$ to $\mathbf{J}_k^{D}$ and 
represents the influence on the signal transmission caused by the existence of target. 
The material-related EM property of the target is implicitly incorporated in the formulation of $\mathbf{X}_k$, which can be leveraged to estimate the material of the target. 
\textcolor{black}{
The matrices $\mathbf{H}_{1,k}$ and $\mathbf{H}_{2,k}$ are also known as the radiation operator matrices and are solely related to the physical laws of EM propagation \cite{operator}. 
Since we assume that the positions of the receiver, the sensing domain $D$, and the transmitter are known, 
$\mathbf{H}_{1,k}$ and $\mathbf{H}_{2,k}$ can be calculated by the computational EM methods 
such as method of moments (MOM) \cite{operator0}, \cite{MOM}, finite element method (FEM) \cite{FEM}, and  finite-difference time-domain (FDTD) \cite{FDTD}. 
}
In this paper, the objective is to retrieve the EM property of the target using the received signals $\tilde{y}_k$ and to further identify the material of the target from $\mathbf{X}_k$.  





\section{Modeling the End-to-End EM Propagation}
In this section, a closed-form expression of $\mathbf{X}_k$ is derived to unfold the influence on the received signals caused by the target scattering process. 
Sensing the target EM property 
is equivalent to reconstructing the difference between the complex relative permittivity of the target and the complex relative permittivity of air at each point $\mathbf{r}^{\prime}$ in domain $D$, denoted as $\chi_k(\mathbf{r}^{\prime})$. 
Since the complex relative permittivity of air is approximately equal to $1$, we can formulate the contrast function as: 
\begin{align}
\chi_k(\mathbf{r}^{\prime})=\epsilon_r(\mathbf{r}^{\prime})+j \sigma(\mathbf{r}^{\prime}) / (\epsilon_0 \omega_k) -1,
\label{def0}  
\end{align}
where $\epsilon_r(\mathbf{r}^{\prime})$ is the real relative permittivity at point $\mathbf{r}^{\prime}$, $\sigma(\mathbf{r}^{\prime})$ is the 
conductivity at point $\mathbf{r}^{\prime}$, $\epsilon_0$ is the vacuum permittivity, and $\omega_k=2\pi f_k$ is the angular frequency of electromagnetic waves of the $k$-th subcarrier. The distributions of $\epsilon_r(\mathbf{r}^{\prime})$ and $\sigma(\mathbf{r}^{\prime})$ are the targets that need to be recovered in the EM property sensing task. 
The total electric field corresponding to the $k$-th subcarrier 
inside the domain $D$ can be expressed using the Lippmann-Schwinger (LS) equation as \cite{lipp}, \cite{lipp2}
\begin{align}
E_{k}^{t} ({\mathbf {r}}) = E_{k}^{i} ({\mathbf {r}})+k_{k}^{2}  \int  \limits _{D} {G_{k}({\mathbf {r},\mathbf{r}}') } \chi_{k} ({\mathbf {r}}') E_{k}^{t} ({ \mathbf {r}}')d{\mathbf {r}}', \mathbf {r}\in D , 
\label{lipp1}%
\end{align}
where $k_k=\frac{2\pi}{\lambda_k}$ is the wavenumber of the $k$-th subcarrier, while
$E_{k}^{t} ({\mathbf {r}})$,       
$E_{k}^{i} ({\mathbf {r}})$, 
and ${G_{k}({\mathbf {r},\mathbf{r}}')}$ denote the total electric field, the incident electric field, and the Green's function, respectively. 
Consider a 2D transverse magnetic (TM) scenario, and ${G_{k}({\mathbf {r},\mathbf{r}}')}$ can then be formulated as \cite{lipp}, \cite{lipp2}
\begin{equation}
{G_{k}({\mathbf {r},\mathbf{r}}')}=\frac{j}{4} H_0^{(2)}\left(k_k\left\Vert\mathbf{r}-\mathbf{r}^{\prime}\right\Vert_2\right) ,
\end{equation}
where $H_0^{(2)}$ is the zero-order Hankel function of the second kind.



Let the vectors $\mathbf{E}_k^{s,D} \in \mathbb{C}^{M \times 1}$, $\mathbf{E}_k^{t,D} \in \mathbb{C}^{M \times 1}$, and $\boldsymbol{\chi}_k \in \mathbb{C}^{M \times 1}$ collect the scattering electric field, the total electric field, and $\chi_k(\mathbf{r})$ at all $M$ sampling points in $D$, respectively.
Denote $\mathbf{G}_k \in \mathbb{C}^{M \times M}$ as the discretized matrix of the integral kernel $k_k^2 G_k\left(\mathbf{r}-\mathbf{r}^{\prime}\right)$ in (\ref{lipp1}) using MOM, where both $\mathbf{r}$ and $\mathbf{r}'$ are on the discretized sampling points within the domain $D$. 
The total electric field in $D$ can be obtained by adding the incident field and the scattering field, which means $\mathbf{E}_k^{t,D}$ can be written in the discretized form of (\ref{lipp1}) as:
\begin{equation}
\mathbf{E}_k^{t,D}=\mathbf{E}_k^{i,D}+\mathbf{E}_k^{s,D}
=\mathbf{E}_k^{i,D}+ \mathbf{G}_k \textrm {diag}\left ({\boldsymbol{\chi }_{k}}\right) \mathbf{E}_k^{t,D}  .
\label{lip3}
\end{equation}



In order to yield a closed-form expression of $\mathbf{E}_k^{t,D}$, we reformulate equation (\ref{lip3}) in the non-linear form with respect to $\textrm {diag}\left ({\boldsymbol{\chi }_{k}}\right)$ as:  
\begin{equation}
\mathbf{E}_k^{t,D}=(\mathbf{I}-\mathbf{G}_k \textrm {diag}\left ({\boldsymbol{\chi }_{k}}\right))^{-1} \mathbf{E}_k^{i,D} .
\label{lip3.5}
\end{equation}



After obtaining $\mathbf{E}_k^{t,D}$,
the equivalent contrast source in $D$ is defined as \cite{lipp}, \cite{lipp2}
\begin{align}
\mathbf{J}_k^{D}={\boldsymbol{\chi}_{k}}\odot\mathbf{E}_k^{t,D}=\textrm {diag}\left ({\boldsymbol{\chi}_{k}}\right)\mathbf{E}_k^{t,D},
\end{align}
which can be regarded as the source of scattering EM waves. 
Considering the radiation from the equivalent contrast source $\textrm {diag}\left ({\boldsymbol{\chi}_{k}}\right)\mathbf{E}_k^{t,D}$ transmitted through the channel $\mathbf{H}_{2,k}$, we can formulate $\tilde{\mathbf{y}}_{k}$ as \cite{m-born}, \cite{v-born}
\begin{align}
\tilde{\mathbf{y}}_{k} 
&= \mathbf{H}_{2,k} \textrm {diag}\left ({\boldsymbol{\chi }_{k}}\right)\mathbf{E}_k^{t,D} + \! \tilde{\mathbf{n}}_k \nonumber \\
&= \mathbf{H}_{2,k}  
\textrm {diag}\left ({\boldsymbol{\chi}_{k}}\right)
(\mathbf{I}-\mathbf{G}_k \textrm {diag}\left ({\boldsymbol{\chi }_{k}}\right))^{-1} \mathbf{E}_k^{i,D}+ \! \tilde{\mathbf{n}}_k.  
\label{eee}
\end{align}
Taking into account $\mathbf{E}_k^{i,D}=\mathbf{H}_{1,k} \mathbf{w}_k$, the received signal transmitted through the target scattering path can be written 
as:
\begin{align}
\tilde{\mathbf{y}}_{k} \! = \mathbf{H}_{2,k}  
\textrm {diag}\left ({\boldsymbol{\chi}_{k}}\right)
\left(\mathbf{I}-\mathbf{G}_k \textrm {diag} \left ({\boldsymbol{\chi }_{k}}\right)\right)^{-1} \mathbf{H}_{1,k} \mathbf{w}_k \! + \! \tilde{\mathbf{n}}_k.
\end{align}
Therefore, the closed-form expression of $\mathbf{X}_k$ in (\ref{X}) can be written as:
\begin{align}
\mathbf{X}_k =  
\textrm {diag}\left ({\boldsymbol{\chi}_{k}}\right)
(\mathbf{I}-\mathbf{G}_k \textrm {diag} \left ({\boldsymbol{\chi }_{k}}\right))^{-1} .  
\label{X_1}
\end{align}
For those sampling points occupied by the air within $D$, the corresponding rows of $\mathbf{X}_k$ are all zeros, and those points are not regarded as EM sources of scattering waves.
Note that (\ref{X_1}) is a general formulation regardless of the positions of the transmitter and the receiver. Although we consider a bistatic sensing scenario in this paper, (\ref{X_1}) can also be used in a monostatic sensing scenario where the transmitter is in close proximity to the receiver.

Denote $\tilde{\mathbf{W}}_k =\left[\mathbf{w}_{k,1}, \mathbf{w}_{k,2}, \cdots, \mathbf{w}_{k,I}\right] \in \mathbb{C}^{N_t \times I}$ as the digital transmitter beamformer stacked by time of the $k$-th subcarrier.
Then the overall received pilot signals $\mathbf{Y}_k\in \mathbb{C}^{N_r \times  I}$ can be formulated in a compact form as:
\begin{align}
\mathbf{Y}_k &=  [\tilde{\mathbf{y}}_{k,1},\tilde{\mathbf{y}}_{k,2},\cdots,\tilde{\mathbf{y}}_{k,I}]
\nonumber \\ 
&=  
\mathbf{H}_{2,k}
\textrm {diag}
\left ({\boldsymbol{\chi}_{k}}\right)
(\mathbf{I}-\mathbf{G}_k \textrm {diag}\left ({\boldsymbol{\chi }_{k}}\right))^{-1} 
\mathbf{H}_{1,k} \nonumber \\ & \times \left[\mathbf{w}_{k,1}, \mathbf{w}_{k,2}, \cdots, \mathbf{w}_{k,I}\right] 
 + \left[\tilde{\mathbf{n}}_{k,1}, \tilde{\mathbf{n}}_{k,2}, \cdots, \tilde{\mathbf{n}}_{k,I}\right]
\nonumber \\
& =  \mathbf{H}_{2,k}
\textrm {diag}
\left ({\boldsymbol{\chi}_{k}}\right)
(\mathbf{I}-\mathbf{G}_k \textrm {diag}\left (\boldsymbol{\chi }_{k}\right))^{-1} 
\mathbf{H}_{1,k} \tilde{\mathbf{W}}_k
 \nonumber \\
& + \left[\tilde{\mathbf{n}}_{k,1}, \tilde{\mathbf{n}}_{k,2}, \cdots, \tilde{\mathbf{n}}_{k,I}\right].    
\label{y1}
\end{align} 

Denote the vectorized noise as $\mathbf{n}_{k}=\left[\tilde{\mathbf{n}}_{k,1}^\top, \tilde{\mathbf{n}}_{k,2}^\top, \cdots, \tilde{\mathbf{n}}_{k,I}^\top\right]^\top$.
Then we can vectorize $\mathbf{Y}_k$ to $\mathbf{y}_k \in \mathbb{C}^{I N_r \times  1}$ as:
\begin{align}
\mathbf{y}_k &= \mathrm{vec}(\mathbf{Y}_k)
\nonumber \\ 
& \overset{(a)}{=}  
\left[\left(\tilde{\mathbf{W}}_k^\top
\mathbf{H}_{1,k}^\top
\left(\mathbf{I}-\mathbf{G}_k \textrm {diag}\left ({\boldsymbol{\chi }_{k}}\right)\right)^{-\top}\right) \circ \mathbf{H}_{2,k} \right] \boldsymbol{\chi }_{k} + 
\mathbf{n}_{k},
\label{n1}
\end{align} 
where $\overset{(a)}{=}$ in (\ref{n1}) comes from the equality: $\mathrm{vec}\left(\mathbf{A} \textrm{diag}(\mathbf{b}) \mathbf{C}\right)=\left(\mathbf{C}^T \circ \mathbf{A}\right) \mathbf{b}$. 
In order to derive a quasi-linear sensing model, we further define the sensing matrix $\mathbf{D}_k \in \mathbb{C}^{I N_r \times M  }$ as: 
\begin{align}
\mathbf{D}_k &\overset{\Delta}{=}
\left(\tilde{\mathbf{W}}_k^\top
\mathbf{H}_{1,k}^\top
\left(\mathbf{I}
-\mathbf{G}_k \textrm {diag}\left ({\boldsymbol{\chi }_{k}}\right)\right)^{-\top}\right) \circ \mathbf{H}_{2,k} .   
\label{sense1}
\end{align}
Then the sensing equation with respect to $\boldsymbol{\chi}_k$ can be formulated as: 
\begin{align}
\mathbf{y}_k & = \mathbf{D}_k \boldsymbol{\chi}_k
+\mathbf{n}_k.
\label{sensing}
\end{align}
Note that although we call (\ref{sensing}) a quasi-linear sensing model, $\mathbf{y}_k$ is actually nonlinear with respect to $\boldsymbol{\chi}_k$ because $\mathbf{D}_k$ also depends on $\boldsymbol{\chi}_k$ and is therefore unknown.

\section{EM Property Sensing and Material Identification}
\subsection{Compressive Sensing Based Method}
In this section, we will introduce the EM property sensing
 and material identification 
 method by fusing received pilot signals of all $K$ subcarriers. The sensing equation (\ref{sensing}) of the $k$-th subcarrier can be written as:
\begin{align}
\mathbf{y}_k=\mathbf{D}_k \boldsymbol{\chi}_k +\mathbf{n}_k=\mathbf{D}_k\left(\boldsymbol{\varepsilon}-\mathbf{1}+\frac{j}{\omega_k \boldsymbol{\varepsilon}_0} \boldsymbol{\sigma}\right) + \mathbf{n}_k .
\label{17}
\end{align}
Denote the central angular frequency as $\omega_c$ and the corresponding contrast vector as $\boldsymbol{\chi}_c = \boldsymbol{\varepsilon}-\mathbf{1}+\frac{j}{\omega_c \boldsymbol{\varepsilon}_0} \boldsymbol{\sigma}$.
Seperating the real and imaginary part of the equation (\ref{17}) and extracting frequency-related components in $\boldsymbol{\chi}_k$, we can derive the non-dimensional equation as: 
\begin{align}
\left[\!\begin{array}{l}
\Re\left(\mathbf{y}_k\right) \\
\Im\left(\mathbf{y}_k\right)
\end{array}\!\right]\!
& =\!\left[\!\begin{array}{cc}
\Re\left(\mathbf{D}_k\right) & -\Im\left(\mathbf{D}_k\right) \\
\Im\left(\mathbf{D}_k\right) & \Re\left(\mathbf{D}_k\right)
\end{array}\!\right]\!\left[\!\begin{array}{cc}
\mathbf{I} & \mathbf{0} \\
\mathbf{0} & \frac{\omega_c}{\omega_k}
\mathbf{I}
\end{array}\!\right]\!
\left[\!\begin{array}{c}
\boldsymbol{\varepsilon}-\mathbf{1} \\
\frac{1}{\omega_c \boldsymbol{\varepsilon}_0}\boldsymbol{\sigma}
\end{array}\!\right]\! \nonumber\\
&+ \left[\begin{array}{l}
\Re\left(\mathbf{n}_k\right) \\
\Im\left(\mathbf{n}_k\right)
\end{array}\right] \nonumber \\
& =   
\left[\!\!\begin{array}{cc}
\Re\left(\mathbf{D}_k\right) & -\frac{\omega_c}{\omega_k}\Im\left(\mathbf{D}_k\right) \\
\Im\left(\mathbf{D}_k\right) & \frac{\omega_c}{\omega_k}\Re\left(\mathbf{D}_k\right)
\end{array}\!\!\right]\!
\left[\!\begin{array}{c}
\boldsymbol{\varepsilon}-\mathbf{1} \\
\frac{1}{\omega_c \boldsymbol{\varepsilon}_0}\boldsymbol{\sigma}
\end{array}\!\!\right]
+ \left[\!\!\begin{array}{l}
\Re\left(\mathbf{n}_k\right) \\
\Im\left(\mathbf{n}_k\right)
\end{array}\!\!\!\right].
\label{Ek}
\end{align}
For notational brevity, we can reformulate (\ref{Ek}) as:
\begin{align}
\mathbf{z}_k=\mathbf{E}_k \mathbf{s} + \left[\begin{array}{l}
\Re\left(\mathbf{n}_k\right) \\
\Im\left(\mathbf{n}_k\right)
\end{array}\right],
\label{Ek2}
\end{align}
where $\mathbf{z}_k = \left[ \Re\left(\mathbf{y}_k\right)^\top , \Im\left(\mathbf{y}_k\right)^\top
\right]^\top$, $\mathbf{s}=[(\boldsymbol{\varepsilon}-\mathbf{1})^\top,(\frac{1}{\omega_c \boldsymbol{\varepsilon}_0}\boldsymbol{\sigma})^\top]^\top$, and $\mathbf{E}_k$ denotes the coefficient matrix in front of $\mathbf{s}$.
We define $\mathbf{s}$ as the EM property vector that needs to be restored. 
Since $\mathbf{s}$ is irrelevant to the subcarrier frequencies, we can stack (\ref{Ek2}) for different subcarriers by column as:
\begin{align}
\left[\begin{array}{c}
\mathbf{z}_1 \\
\vdots \\
\mathbf{z}_K
\end{array}\right]=\left[\begin{array}{c}
\mathbf{E}_1 
\\ 
\vdots \\
\mathbf{E}_K
\end{array}\right] \mathbf{s} 
+
\left[\begin{array}{c}
\Re\left(\mathbf{n}_1\right) \\
\Im\left(\mathbf{n}_1\right) \\
\vdots \\
\Re\left(\mathbf{n}_K\right) \\
\Im\left(\mathbf{n}_K\right) \\
\end{array}\right].
\label{Ek3}
\end{align}
Define the general vector of measurement and the general sensing matrix as:
\begin{align}
\tilde{\mathbf{z}} &= \left[\mathbf{z}_1^\top, \cdots, \mathbf{z}_K^\top \right]^\top ,  \\  
\tilde{\mathbf{E}} &= \left[\mathbf{E}_1^\top, \cdots, \mathbf{E}_K^\top \right]^\top.  \label{assemble}
\end{align}
Since the target typically 
occupies only a small fraction of space in the detection domain $D$, most of the elements of $\mathbf{s}$ are $0$ corresponding to the EM property of the air.  
Moreover, $(\boldsymbol{\varepsilon}-\mathbf{1})$ and $\frac{1}{\omega_c \boldsymbol{\varepsilon}_0}\boldsymbol{\sigma}$ 
share the same support set.
Thus, if $\tilde{K}$ sampling points of $D$ are occupied by the target, then 
$(\boldsymbol{\varepsilon}-\mathbf{1})$ and $\frac{1}{\omega_c \boldsymbol{\varepsilon}_0}\boldsymbol{\sigma}$ will be  $\tilde{K}$-sparse, and 
 $\mathbf{s}$ will be jointly $\tilde{K}$-sparse. 
According to the compressive sensing techniques, $\mathbf{s}$ can be reconstructed as the solution to the following mixed ($\ell1$,$\ell2$)-norm minimization problem 
\begin{subequations}
\begin{align}
\min \  &\|\mathbf{s}\|_{1,2} \\
\text { s.t. }&\|
\tilde{\mathbf{z}}-\tilde{\mathbf{E}} \mathbf{s}\|_2 \leqslant \varepsilon^{\prime}, \quad \mathbf{s} \geq \mathbf{0},
\end{align}
\label{com}%
\end{subequations}
where $\varepsilon^{\prime}$ represents the predefined noise level that should be chosen appropriately. 
The constraint $\mathbf{s} \geq \mathbf{0}$ means all elements of $\mathbf{s}$ are nonnegative and holds  based on the fact that the relative permittivity is not less than $1$ and the conductivity is nonnegative. 
Moreover, $\|\mathbf{s}\|_{1,2}$ is the mixed ($\ell1$,$\ell2$)-norm defined as:
\begin{equation} 
\| \mathbf{s}\|_{1,2}
\overset{\Delta}{=}
{\sum _{m=1}^{M}
\sqrt{s_m^2+s_{m+M}^2}} .    
\end{equation}
The mixed ($\ell1$,$\ell2$)-norm tends to guarantee the joint sparsity of $\mathbf{s}$, because $\boldsymbol{\epsilon}-\mathbf{1}$ and $\boldsymbol{\sigma}$ have the same support set. 
Based on the generalized multiple measurement vector (GMMV) model, the key point in (\ref{com}) is to utilize the joint sparsity structure to improve the sensing ability.
Moreover, (\ref{com}) is a basis pursuit denoising (BPDN) problem that can 
be effectively solved by CVX using spectral projected gradient (SPG) method due to its convexity. 
The order of computational complexity is $\mathcal{O}(\tilde{K}^3 M^3)$ according to \cite{cs2}.



\subsection{Iterative EM Property Sensing Formulation}
Since $\tilde{\mathbf{E}}$ intrinsically depends on $\mathbf{s}$,   
we need to update $\tilde{\mathbf{E}}$ after solving (\ref{com}) and update $\mathbf{s}$ iteratively. 
Let the superscript denote the number of iterations. 
For the initial iteration step, the Born approximation (BA) $\mathbf{E}_D^{t} \approx \mathbf{E}_D^{i}$ can be applied \cite{born,born2,lipp}.
BA has been proven to be accurate if the scattering field is relatively small compared to the incident field on the scatterer and thus can be used to calculate the initial guess \cite{born}.  
Using BA, (\ref{X_1}) reduces to $\textbf{X}_k \approx \mathrm{diag}(\boldsymbol{\chi}_k) $.
Following the same derivation in (\ref{y1}) and (\ref{n1}), 
we can compute the initial sensing matrix defined in (\ref{sense1}) 
under BA  as:  
\begin{align}
\mathbf{D}_k^0 =
\left(\tilde{\mathbf{W}}_k^\top
\mathbf{H}_{1,k}^\top
\right) \circ \mathbf{H}_{2,k} .
\end{align}
Then, $\mathbf{D}_k^0$
is used to compute $\mathbf{E}_k^0$ in (\ref{Ek2}),  and $\mathbf{E}_k^0$ of all subcarriers are assembled into $\tilde{\mathbf{E}}^0$ in (\ref{assemble}). By solving (\ref{com}) with $\tilde{\mathbf{E}}^0$, we can further calculate $\mathbf{s}^0$. 
With the initial guess $\mathbf{s}^0$, the proposed iteration procedure of the inversion algorithm is as follows:

Step 1: Calculate $\boldsymbol{\chi}_k^n$ for all subcarriers using $\mathbf{s}^n$ according to $(\ref{def0})$.

Step 2: Calculate $\mathbf{D}_k^n$ using $\boldsymbol{\chi}_k^n$ for all subcarriers according to $(\ref{sense1})$.

Step 3: Calculate $\mathbf{E}_k^n$ using $\mathbf{D}_k^n$ for all subcarriers according to $(\ref{Ek})$ and $(\ref{Ek2})$ and then assemble $\mathbf{E}_k^n$ for all $k$ into $\tilde{\mathbf{E}}^n$ according to $(\ref{assemble})$.

Step 4: Calculate $\mathbf{s}^{n+1}$  using $\tilde{\mathbf{E}}^n$ according to (\ref{com}).

Step 5: If the predetermined
convergence criterion is satisfied, then STOP. Otherwise, go to Step 1.

It is important to note that except for Step 4, 
the remaining steps do not involve inverse operations and only include linear operations. 
Thus, the algorithm is based on recursive linear approximation to cope with the non-linearity of the EM property sensing problem in (\ref{sense1}) and (\ref{sensing}).

Next, we analyze the computational complexity.  
In each iteration, the computational complexity of Step 1 is $\mathcal{O}(K M)$;
the computational complexity of Step 2 is $\mathcal{O}([M^3+IN_tM+IM^2+IN_rM]K)$;
the computational complexity of Step 3 is $\mathcal{O}(N_r M K)$;
the computational complexity of Step 4 is $\mathcal{O}(\tilde{K}^3 M^3)$.
By combining them, we finally have the overall computational complexity as 
$\mathcal{O}([(M^2+IN_t+IM+IN_r)MK+\tilde{K}^3 M^3] N_{iter})$, where $N_{iter}$ represents the number of iterations for convergence.

\subsection{Material Identification Methodology}
Suppose that the object is known to be constituted by one of several possible materials, whose permittivity and conductivity are measured precisely in advance.   
Note that only materials with obvious differences in permittivity or conductivity can be distinguished. 
The material identification methodology consists of two steps: first clustering and then classification.

In order to determine the material of the target, we first need to distinguish between the part of domain $D$ occupied by the air and the part occupied by the target.      
To accomplish this, we utilize the K-means clustering algorithm to divide the sampling points in $D$ into two categories \cite{kmeans}. 
K-means is an unsupervised algorithm with strong generalization of clusters for different shapes and sizes, and is guaranteed to converge regardless of the target.
Since the relative permittivity and conductivity have different dimensions, we adopt the dimensionless and scale-invariant Mahalanobis distance in the K-means clustering algorithm \cite{mahalanobis}.
The number of clusters is predetermines to be $2$, representing the air and the target respectively. 
The cluster centroid of the air, representing the average permittivity and conductivity values of the air, is expected to be close to the $(1,0)$ point. On the other hand, the cluster centroid of the target, representing its average permittivity and conductivity, is expected to be notably far away from the $(1,0)$ point.

After clustering is performed, the next step is to determine the material category of the target.
This is done by calculating the Mahalanobis distance between the cluster centroid of the target material and the ground-truth values of the permittivity and conductivity for each possible material.
The target is then classified into the material category with the shortest Mahalanobis distance which indicates the closest match between the measured EM properties of the target and the known properties of the materials.

The computational complexity of the material identification is dominated by the K-means clustering algorithm, whose complexity is 
$\mathcal{O}(M I_{iter})$ \cite{kmeans_complexity}, where $I_{iter}$ represents the number of iterations for convergence.

\section{Beamforming Matrix Design Methodology} 
In this section, the beamforming matrix is designed to optimize the initial sensing matrix $\tilde{\mathbf{E}}^0$ based on BA. 
The beamforming matrix design is carried out before the sensing iteration, and is unrelated to the target RPCD distribution.
For notational brevity, we will omit the superscript $0$ in the remainder of this section.

\subsection{Mutual Coherence Based Design Criterion}
The restricted isometry property (RIP) \cite{RIP} and
the mutual coherence \cite{adaptive} of the sensing matrix $\tilde{\mathbf{E}}$ 
are the two most common criteria to evaluate the accuracy of recovery in  (\ref{com}).
The RIP provides the tighter 
bound, while it is NP-hard to evaluate \cite{compressed_sensing}.
The mutual coherence $\mu$ of $\tilde{\mathbf{E}}$ is easier to calculate and is defined as the maximum of $\mu_{ij}(\tilde{\mathbf{E}})$ for 
$i \neq j$ defined as:
\begin{equation}
\mu_{i j} (\tilde{\mathbf{E}})=\frac{\left|\tilde{\mathbf{E}}[:,i]^\top \tilde{\mathbf{E}}[:,j]\right|}{\left\|\tilde{\mathbf{E}}[:,i]\right\|_2\left\|\tilde{\mathbf{E}}[:,j]\right\|_2}=\frac{\left|G_{i j}\right|}{\sqrt{\left|G_{i i}\right|\left|G_{j j}\right|}},
\label{mutual}
\end{equation}
where $G_{i j}$ denotes the $(i,j)$th element of the Gram matrix $\tilde{\mathbf{E}}^\top \tilde{\mathbf{E}}$.


The Gram matrix of the general sensing matrix can be decomposed into the weighted sum of the Gram matrix of the sensing matrix for all $K$ subcarriers according to (\ref{assemble}):
\begin{align}
\tilde{\mathbf{E}}^\top \tilde{\mathbf{E}}
& = \sum_{k=1}^K \mathbf{E}_k^\top \mathbf{E}_k \nonumber \\
& = \sum_{k=1}^K \left[\!\!\!\begin{array}{ll}
\Re\left(\mathbf{D}_k\right) & \!-\frac{\omega_c}{\omega_k}\Im\left(\mathbf{D}_k\right) \\
\Im\left(\mathbf{D}_k\right) & \!\frac{\omega_c}{\omega_k}\Re\left(\mathbf{D}_k\right)
\end{array}\!\!\!\!\right] ^ \top
\!\!\!\left[\!\!\!\begin{array}{ll}
\Re\left(\mathbf{D}_k\right) & \!-\frac{\omega_c}{\omega_k}\Im\left(\mathbf{D}_k\right) \\  
\Im\left(\mathbf{D}_k\right) & \!\frac{\omega_c}  {\omega_k} \Re\left(\mathbf{D}_k\right)  
\end{array}\!\!\!\right] \nonumber \\  
& \overset{(a)}{=} \sum_{k=1}^K
\left[\begin{array}{cc}
\Re (\mathbf{D}_k^H \mathbf{D}_k) & -(\frac{\omega_c}{\omega_k})\Im (\mathbf{D}_k^H \mathbf{D}_k) \\
(\frac{\omega_c}{\omega_k}) \Im (\mathbf{D}_k^H \mathbf{D}_k) & (\frac{\omega_c}{\omega_k})^{2 } \Re (\mathbf{D}_k^H \mathbf{D}_k)
\end{array}\!\right],\! 
\label{trans1}
\end{align}            
where $\overset{(a)}{=}$ comes from the following equality:
\begin{align}
\mathbf{D}_k^H \mathbf{D}_k &= \Re (\mathbf{D}_k)^\top \Re (\mathbf{D}_k) + \Im (\mathbf{D}_k)^\top \Im (\mathbf{D}_k) \nonumber \\
&+j \left[ \Re (\mathbf{D}_k)^\top \Im (\mathbf{D}_k) - \Im (\mathbf{D}_k)^\top \Re (\mathbf{D}_k) \right].
\end{align}
In order to minimize the mutual coherence of the general sensing matrix $\tilde{\mathbf{E}}$, we need to minimize the absolute value of off-diagonal elements in  $\tilde{\mathbf{E}}^\top \tilde{\mathbf{E}}$.
According to (\ref{trans1}), the objective is to 
minimize the absolute value of real parts of the off-diagonal elements in 
$\mathbf{D}_k^H \mathbf{D}_k$ and the absolute value of imaginary parts of all elements in 
$\mathbf{D}_k^H \mathbf{D}_k$ for all $k$.
Since the diagonal elements in 
$\mathbf{D}_k^H \mathbf{D}_k$ are all real numbers, the objective is to minimize the absolute value of both real and imaginary parts of the off-diagonal elements, i.e, to minimize the mutual coherence of $\mathbf{D}_k$ for all $k$.





Substituting the BA-based initial guess $ \mathbf{D}_k^0 =
\left(\tilde{\mathbf{W}}_k^\top
\mathbf{H}_{1,k}^\top
\right) \circ \mathbf{H}_{2,k} $ into $\mathbf{D}_k^H \mathbf{D}_k$, we can compute the $(i,j)$th element of $\mathbf{D}_k^H \mathbf{D}_k$ as (\ref{trans2}) at the top of the next page. 
\begin{figure*}[t]
\begin{align}
\left(\mathbf{D}_k^H \mathbf{D}_k \right) [i,j]
&=\left(\left(\tilde{\mathbf{W}}_k^\top
\mathbf{H}_{1,k}^\top
\right)[:,i] \otimes 
\mathbf{H}_{2,k}[:,i]\right)^H
\left(\left(\tilde{\mathbf{W}}_k^\top
\mathbf{H}_{1,k}^\top
\right)[:,j] \otimes 
\mathbf{H}_{2,k}[:,j]\right) \nonumber \\
& = \left(\left(\tilde{\mathbf{W}}_k^\top
\mathbf{H}_{1,k}^\top
\right)[:,i]^H \otimes 
\mathbf{H}_{2,k}[:,i]^H \right)
\left(\left(\tilde{\mathbf{W}}_k^\top
\mathbf{H}_{1,k}^\top
\right)[:,j] \otimes 
\mathbf{H}_{2,k}[:,j]\right) \nonumber \\
& \overset{(a)}{=} \left( \left(\tilde{\mathbf{W}}_k^\top
\mathbf{H}_{1,k}^\top
\right)[:,i]^H \left(\tilde{\mathbf{W}}_k^\top
\mathbf{H}_{1,k}^\top
\right)[:,j] \right) \otimes  
\left(
\mathbf{H}_{2,k}[:,i]^H \mathbf{H}_{2,k}[:,j] \right)   \nonumber \\
& \overset{(b)}{=}  \left( \left(\tilde{\mathbf{W}}_k^\top
\mathbf{H}_{1,k}^\top
\right)[:,i]^H \left(\tilde{\mathbf{W}}_k^\top
\mathbf{H}_{1,k}^\top
\right)[:,j] \right) 
\left(
\mathbf{H}_{2,k}[:,i]^H \mathbf{H}_{2,k},[:,j] \right) 
\label{trans2}
\end{align}
\hrulefill
\vspace*{4pt}
\end{figure*}
In (\ref{trans2}), $\overset{(a)}{=}$ comes from the equality $(\mathbf{A} \otimes \mathbf{B})(\mathbf{C} \otimes \mathbf{D})=(\mathbf{A} \mathbf{C}) \otimes (\mathbf{B} \mathbf{D})$, 
while $\overset{(b)}{=}$ holds true because the Kronecker product is taken between two scalars and is equivalent to the scalar product.
Integrating (\ref{trans2}) for all elements in $\mathbf{D}_k^H \mathbf{D}_k$, we can formulate 
$\mathbf{D}_k^H \mathbf{D}_k$ as: 
\begin{align}
\!\mathbf{D}_k^H \mathbf{D}_k \! = \! \left( \left(\tilde{\mathbf{W}}_k^\top
\mathbf{H}_{1,k}^\top      
\right)^H \left(\tilde{\mathbf{W}}_k^\top
\mathbf{H}_{1,k}^\top  
\right) \right)\! \odot \!\left( 
\mathbf{H}_{2,k}^H \mathbf{H}_{2,k}
\right).
\label{hada1}
\end{align}

\begin{figure*}[t]
  \centering
\begin{minipage}[t]{0.32\linewidth}
\subfigure[$d=10$ m]{
\includegraphics[width=6 cm,height=5.5cm]{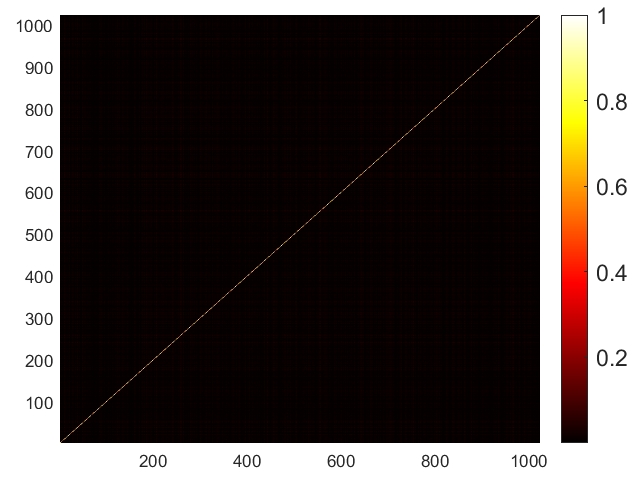}} 
\end{minipage}
\begin{minipage}[t]{0.32\linewidth}
\subfigure[$d=20$ m]{
\includegraphics[width=6 cm,height=5.5cm]{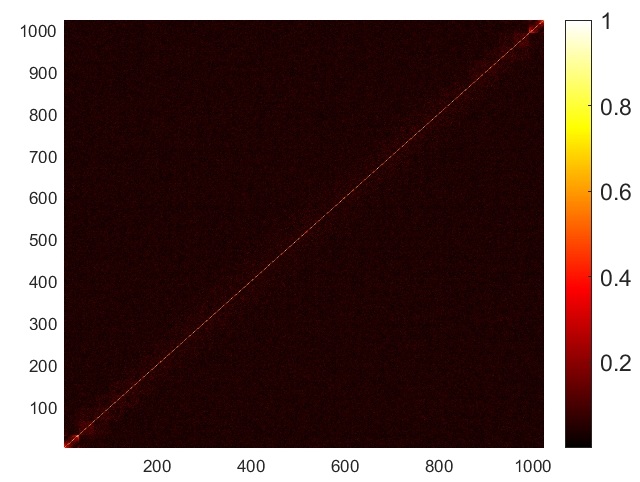}}  
\end{minipage}
\begin{minipage}[t]{0.32\linewidth}
\subfigure[$d=40$ m]{  
\includegraphics[width=6 cm,height=5.5cm]{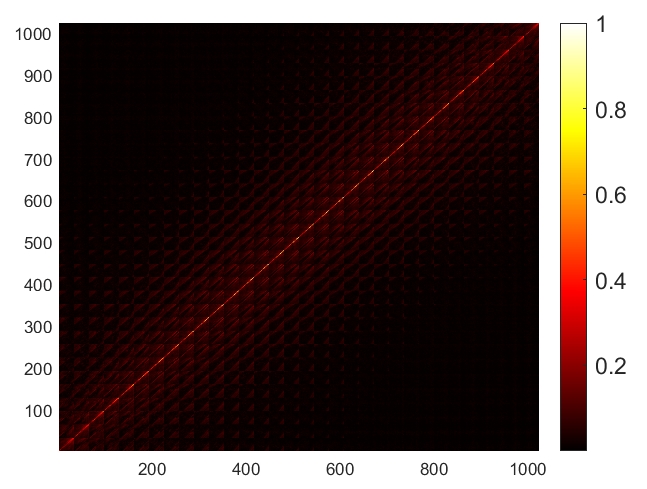}}   
\end{minipage} 
\caption{ Absolute values of Gram matrix $\tilde{\mathbf{E}}^\top \tilde{\mathbf{E}}$ normalized by its largest element is demonstrated  
after designing the beamforming matrix for different $d$.  
}
\label{gram1}
\end{figure*}

Since $\mathbf{H}_{2,k}^H \mathbf{H}_{2,k}$ is determined by the positions of the target and the receiver, 
$\mathbf{H}_{2,k}^H \mathbf{H}_{2,k}$ is not subject to change.
Thus, minimizing the off-diagonal elements of $\mathbf{D}_k^H \mathbf{D}_k$ is equivalent to minimizing the off-diagonal elements of $\left(\tilde{\mathbf{W}}_k^\top
\mathbf{H}_{1,k}^\top      
\right)^H \left(\tilde{\mathbf{W}}_k^\top
\mathbf{H}_{1,k}^\top  
\right)$.           
Hence, we focus on designing $\left(\tilde{\mathbf{W}}_k^\top
\mathbf{H}_{1,k}^\top      
\right)^H \left(\tilde{\mathbf{W}}_k^\top
\mathbf{H}_{1,k}^\top  
\right)$ in (\ref{hada1}) to minimize the mutual coherence of $ \mathbf{D}_k$.    


Since we consider a fully-digital beamforming transmitter,    
$\mathbf{D}_k$ can be designed in the identical way for any subcarrier of an arbitrary index $k$.
Thus, we will omit the subscript $k$ 
for notational brevity in the remainder of this section.



Denote 
$\mathbf{D}_p \overset{\Delta}{=}
\mathbf{H}_1 ^ \top $ and $\mathbf{W} \overset{\Delta}{=} 
\tilde{\mathbf{W}}^{\top}$. 
In order to minimize 
the mutual coherence of $ \mathbf{W} \mathbf{D}_p$, we can solve the following optimization problem that is cast in terms of the Gram matrix $\mathbf{G}_p = (\mathbf{W} \mathbf{D}_p)^H(\mathbf{W} \mathbf{D}_p) \in \mathbb{C}^{M \times M}$:
\begin{subequations}
\begin{align} 
\hat{\mathbf{W}} = & \arg \, \min _{\mathbf{W }, \, \alpha} \, \Vert  (\mathbf{W D}_{p})^{H} (\mathbf{W D}_{p})- \alpha \mathbf{T} \Vert _{F} ^{2} \label{Za} \\ \mathrm{s.t.\,\,} 
& \mathrm{tr}(\mathbf{W }^H \mathbf{W }) \le \frac{P}{K} ,
\end{align}
\label{p0}%
\end{subequations}
where $\mathbf{T} \in \mathbb{C}^{M \times M}$ denotes the target matrix of $\mathbf{G}_{p}$,
$P$ denotes the general power budget at the transmitter which is assumed to be evenly allocated to all $K$ subcarriers, and $\alpha$ denotes the auxiliary scaling factor.  
The mutual coherence of $\mathbf{W} \mathbf{D}_p$ is not affected by the scaling factor that is utilized to satisfy the power constraint.
We refer to equation (\ref{p0}) as the mutual coherence based design criterion due to its connection with the mutual coherence outlined in (\ref{mutual}). This relationship is further exemplified by the interplay among various Gram matrices $\tilde{\mathbf{E}}^\top \tilde{\mathbf{E}}$, ${\mathbf{D}_k}^H {\mathbf{D}_k}$, and $(\mathbf{W} \mathbf{D}_p)^H(\mathbf{W} \mathbf{D}_p)$, as detailed in equations (\ref{hada1}) and (\ref{trans1}). 




In order to build a 
robust sensing system that is able to deal with the representation error, a promising approach is to employ the Gram matrix of $\mathbf{D}_p$ as the target Gram \cite{adaptive}, \cite{adaptive2}.  
Similar to \cite{adaptive}, $\mathbf{T}$ is judiciously  
designed using the element-wise absolute value of $\mathbf{D}_p^H \mathbf{D}_p$, where the element of $\mathbf{T}$ is formulated as:
\begin{equation} 
T_{ij} = \left|\mathbf{D}_p[:,i]^H \mathbf{D}_p[:,j] \right|, 1\leq i,j \leq M .
\end{equation}







Since multiplying $\mathbf{W}$ by a scalar does not change the mutual coherence of $\mathbf{W} \mathbf{D}_p$, we can remove the power budget constraint and neglect the scaling factor $\alpha$ in (\ref{p0}) to find an unnormalized 
beamforming matrix $\mathbf{W}^u$ without the power constraint.
Thus, (\ref{p0}) can be transformed into an unconstrained optimization problem over $\mathbf{W}^u$ as: 
\begin{align} 
\hat{\mathbf{W}}^u = & \arg \, \min _{\mathbf{W }^u} \Vert (\mathbf{W}^u \textbf{D}_{p})^{H} (\mathbf{W}^u \textbf{D}_{p})-  \mathbf{T} \Vert _{F} ^{2} .
\label{p1}
\end{align}

Let the singular value decomposition (SVD) of $\mathbf{\mathbf{D}_p}$ be 
$\mathbf{\mathbf{D}_p}=\mathbf{U}_{\mathbf{D}_p}
\boldsymbol{\Sigma}_{\mathbf{D}_p}
\mathbf{V}_{\mathbf{D}_p}^H$. 
By defining
$\mathbf{\Psi }=
\mathbf{W }^u \mathbf{U}_{\mathbf{D}_p} \boldsymbol{\Sigma}_{\mathbf{D}_p}$, (\ref{p0}) can be written as:
\begin{equation} \hat{\mathbf{\Psi }} = \arg \, \min _{\mathbf{\Psi }} \Vert \mathbf{V}_{\mathbf{D}_p} \mathbf{\Psi }^{H} \mathbf{\Psi } \mathbf{V}_{\mathbf{D}_p}^{H} - \mathbf{T} \Vert _{F} ^{2}  . 
\label{p2}
\end{equation}
Since $\mathbf{V}_{\mathbf{D}_p}$ is a unitary matrix that does not change the Frobenius norm by multiplication, (\ref{p2}) is equivalent to
\begin{align} 
\hat{\mathbf{\Psi }} = \arg \, \min _{\mathbf{\Psi }} \Vert \mathbf{ \Psi }^{H} \mathbf{ \Psi }-\mathbf{V}_{\mathbf{D}_p}^{H} \mathbf{T} \mathbf{V}_{\mathbf{D}_p} \Vert _{F} ^{2}.
\label{quartic}
\end{align}
Since (\ref{quartic}) is a quartic optimization problem that is neither convex nor concave over $\mathbf{\Psi}$, problem (\ref{quartic}) is non-convex.
Therefore, we recast (\ref{quartic}) in terms of the Gram matrix $\mathbf{G}_{\Psi}=\boldsymbol{\Psi}^H \boldsymbol{\Psi}$ and the new target matrix $\mathbf{Z}=\mathbf{V}_{\mathbf{D}_p}^H \mathbf{T} 
\mathbf{V}_{\mathbf{D}_p}$ as:
\begin{subequations}
\begin{align} 
\hat{\mathbf{G}}_\Psi = & \arg \, \min _{\mathbf{G}_{\Psi }} \Vert \mathbf{G}_\Psi -\mathbf{Z} \Vert _{F} ^{2} \label{conv1} \\ 
\mathrm{s.t.\,\,} & \mathbf{G}_\Psi \succeq \textbf{0}, \, \text{rank} (\mathbf{G}_\Psi) \leq I.
\label{rank}
\end{align}
\label{p3}%
\end{subequations}
The rank constraint in (\ref{rank}) comes from the fact that the dimensions of $\boldsymbol{\Psi}$ are $I \times N_t$, which determines the rank of $\mathbf{G}_\Psi \in \mathbb{C}^{N_t \times N_t}$ to be smaller than or equal to $I$.
Although (\ref{conv1}) is convex over $\mathbf{G}_{\Psi }$ after transformation, 
the rank constraint in (\ref{rank}) makes (\ref{p3}) non-convex.
Note that we do not employ semidefinite relaxation (SDR) \cite{SDR}, because the optimization problem in (\ref{p3}) is a semi-definite low-rank approximation problem, 
whose optimal solution can be obtained using the generalized Eckart-Young Theorem as \cite{low_rank_theorem}, \cite{low_rank}: 
\begin{equation} \hat{\mathbf{G}}_\Psi = \sum_{i=1}^{\text{min} \lbrace I , z \rbrace } \lambda _{i} \mathbf{q}_{i} \mathbf{q}_{i}^{H}, 
\label{p4}
\end{equation}
where the Hermitian target matrix $\mathbf{Z}$ has the eigen-decomposition: $\mathbf{Z}=\mathbf{Q} \mathbf{\Lambda} \mathbf{Q}^H=\sum_{i=1}^M \lambda_i \mathbf{q}_i \mathbf{q}_i^H$ with $\lambda_1 \geq \ldots \geq \lambda_M$, and $z$ denotes the number of non-negative eigenvalues of $\mathbf{Z}$. After the optimal positive semi-definite matrix $\hat{\mathbf{G}}_{\Psi}$ is obtained by (\ref{p4}), $\hat{\boldsymbol{\Psi}}$ that satisfies $\hat{\boldsymbol{\Psi}}^H \hat{\boldsymbol{\Psi}}=\hat{\mathbf{G}}_{\Psi}$ is then found from the eigen-decomposition of $\hat{\mathbf{G}}_{\Psi}$:
\begin{align}
\hat{\boldsymbol{\Psi}}= \begin{cases}\mathbf{\Lambda}_I \mathbf{Q}_I^H, 
& \text { if } I \leq z \\ 
{\left[\mathbf{Q}_z \mathbf{\Lambda}_z^H , \mathbf{0}\right]^H,} & \text { if } I > z\end{cases} ,
\end{align}
where $\mathbf{Q}_I$ (or $\mathbf{Q}_z$) is defined as the matrix that only takes the first $I$ (or $z$) columns of $\mathbf{Q}$, and $\mathbf{\Lambda}_I$ (or $\mathbf{\Lambda}_z$) is defined as the diagonal matrix that only takes the first $I$ (or $z$) columns and rows of $\mathbf{\Lambda}$. 
Then, the optimal unnormalized beamforming matrix $\hat{\mathbf{W}}^u$ that solves (\ref{p1}) can be formulated as:
\begin{equation}
\hat{\mathbf{W}}^u
=\hat{\boldsymbol{\Psi}} \boldsymbol{\Sigma}_{\mathbf{D}_p}^{-1} \mathbf{U}_{\mathbf{D}_p}^H .
\end{equation}
Considering that the power budget $P$ is evenly allocated to $K$ subcarriers at the transmitter, the optimal beamforming matrix should be normalized as: 
\begin{equation}
\hat{\mathbf{W}} = \sqrt{\frac{P}{K}} \frac{\hat{\mathbf{W}}^u}{\Vert\hat{\mathbf{W}}^u\Vert_F}.
\end{equation}



\subsection{Computational Complexity}
In this subsection, we analyze the computational complexity of designing the beamforming matrix $\hat{\mathbf{W}}$ in the previous subsection.
For each subcarrier, the computational complexity of calculating SVD of $\mathbf{D}_p$ is $\mathcal{O}(\min(N_t^2 M,N_t M^2))$;
the computational complexity of calculating  $\mathbf{T}$ and $\mathbf{Z}$ is $\mathcal{O}( M^2 N_t + N_t^2 M)$;
the computational complexity of calculating  $\hat{\mathbf{G}}_{\boldsymbol{\Psi}}$ is $\mathcal{O}(\min(I,z) M^2)$;
the computational complexity of calculating  $\hat{\boldsymbol{\Psi}}$ and $\hat{\mathbf{W}}^u$ is $\mathcal{O}(\min(I,z) N_t^2 )$;
the computational complexity of calculating $\hat{\mathbf{W}}$ is $\mathcal{O}(I N_t)$.
Thus, the overall computational complexity of designing the beamforming matrices $\hat{\mathbf{W}}$ of all $K$ subcarriers 
is given by $\mathcal{O}( [M^2 N_t + N_t^2 M+\min(I,z)(M^2+ N_t^2)]K )$.


\subsection{Influence of Transmitter-Target Distance on Sensing}
In this subsection, we analyze the influence of the transmitter-target distance based on the effective degrees of freedom (EDOF) of the transmitter-target sensing system \cite{myEDOF}.
Define the correlation matrix of the transmitter-target channel of an arbitrary subcarrier as $\mathbf{R}=\mathbf{H}_1 \mathbf{H}_1^H$.
The EDOF of the transmitter-target sensing system is a function of $\mathbf{R}$ denoted by $\Xi$, and can be approximately calculated as \cite{1}, \cite{2}: 
\begin{equation}
\Xi = \left(\frac{\operatorname{tr}(\mathbf{R})}{\|\mathbf{R}\|_{F}}\right)^{2}=\frac{\left(\sum_{i=1}^M \sigma_{i}\right)^{2}}{\sum_{i=1}^M \sigma_{i}^{2}},
\label{dis}   
\end{equation}  
where $\sigma_{i}$ is the $i$-th eigenvalue of $\mathbf{R}$. 
In the far-field sensing scenarios, the leading eigenvalue of $\mathbf{R}$ is significantly larger than the other eigenvalues, and the EDOF is close to $1$ corresponding to the only
dominant sensing mode where a planar wave travels from the transmitter to the target \cite{9}. 
Thus, the target is required to be located within the near field of the transmitter where the EDOF is significantly larger than $1$, which enables accurate RPCD reconstruction and material identification.


\section{Simulation Results and Analysis}
\begin{figure}[t]
  \centering
\centerline{\includegraphics[width=8.4cm,height=6.5cm]{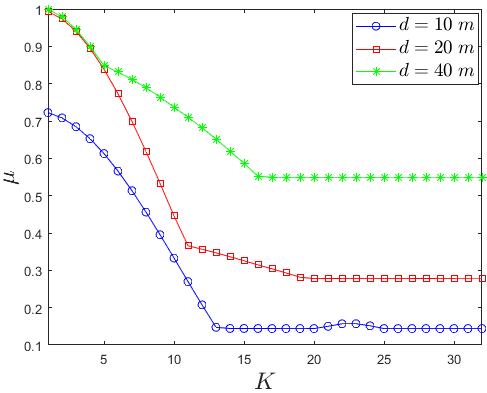}}  \caption{\textcolor{black}{Mutual coherence of the general sensing matrix $\tilde{\mathbf{E}}$ with the increase of the number of subcarriers $K$.}}
  \label{mut}
\end{figure}
In this section, 
we use the EM simulation software Ansys HFSS to simulate the transmission process in the ISAC system and to generate the received signals, which are used in reconstructing the EM property and identifying the material of the target.   
Suppose the center of the target is located at the origin.  
We consider a uniform linear array (ULA) transmitter equipped with $N_t=1024$ antennas.
The ULA transmitter is parallel to the $y$-axis, and its center is located at $(x_t,y_t) = (d,0)$~m, where $d$ denotes the distance from the transmitter to the target.
Suppose a ULA receiver equipped with $N_r=16$ antennas 
is located at $(x_r,y_r) = (-20,-20)$~m and is parallel to the ULA transmitter.
The inter-antenna spacing for both the transmitter and the receiver is set as $\lambda_c/2$. 
The central frequency is $f_c = 30$ GHz and the frequency step of OFDM subcarriers is $\Delta f = 200$ KHz.
In each subcarrier, $I=32$ pilot OFDM symbols are transmitted. 
Suppose prior knowledge determines that the target is located within the region $D = [-1,1] \times [-1,1]$~$\mathrm{m}^2$.
We choose the number of sampling points in the domain $D$ as $ M = 32 \times 32 = 1024$. 
Each sampling point in $D$ can be regarded as a pixel in the image, i.e., the RPCD image is composed of $32 \times 32 = 1024$ pixels.
More pixels could hardly improve the RPCD reconstruction quality due to the resolution limit.


Besides, we introduce the normalized mean square error (NMSE) of RPCD reconstruction as the criterion to quantitatively describe the performance of EM property sensing  
\begin{align}
\mathrm{NMSE}&=
10 \log_{10} 
\frac{\left\|\mathbf{s}-\hat{\mathbf{s}}\right\|_2^{2}}{\left\|\mathbf{s}\right\|_2^{2}} 
\nonumber\\
&= 
10 \log_{10}
\frac{\left\|\boldsymbol{\epsilon}_r-\hat{\boldsymbol{\epsilon}}_r\right\|_2^{2} + \frac{1}{\omega_c^2 \epsilon_0^2}\left\|\boldsymbol{\sigma}-\hat{\boldsymbol{\sigma}}\right\|_2^{2}} {\left\|\boldsymbol{\epsilon}_r\right\|_2^{2} + \frac{1}{\omega_c^2 \epsilon_0^2}\left\|\boldsymbol{\sigma}\right\|_2^{2}}.
\label{NMSE}
\end{align}
According to (\ref{NMSE}), the NMSE of RPCD combines the estimation error of both $\boldsymbol{\epsilon}$ and $\boldsymbol{\sigma}$, which comprehensively assess the EM property sensing performance.

In order to provide a visual example of the mutual coherence of the general sensing matrix $\tilde{\mathbf{E}}$,
we show the absolute values of Gram matrix $\tilde{\mathbf{E}}^\top \tilde{\mathbf{E}}$ normalized by its largest element in Fig.~\ref{gram1}.
We set the number of subcarriers to be $K=32$.
It is seen from Fig.~\ref{gram1} that the off-diagonal components are generally much smaller than the diagonal components, indicating that the mutual coherence is relatively small after designing the beamforming matrix. 
However, with an increase in the distance between the transmitting antennas and the target, the mutual coherence also increases correspondingly.
This is because as the distance between the transmitting antennas and the target increases, the channel matrix can become more ill-conditioned or closer to being singular. 
The correlation degree of adjacent lattice points in $D$ is then increased, which will be reflected in the reduction of the reconstructed RPCD resolution.

\begin{figure}[t]
  \centering
\centerline{\includegraphics[width=8.4cm,height=6.5cm]{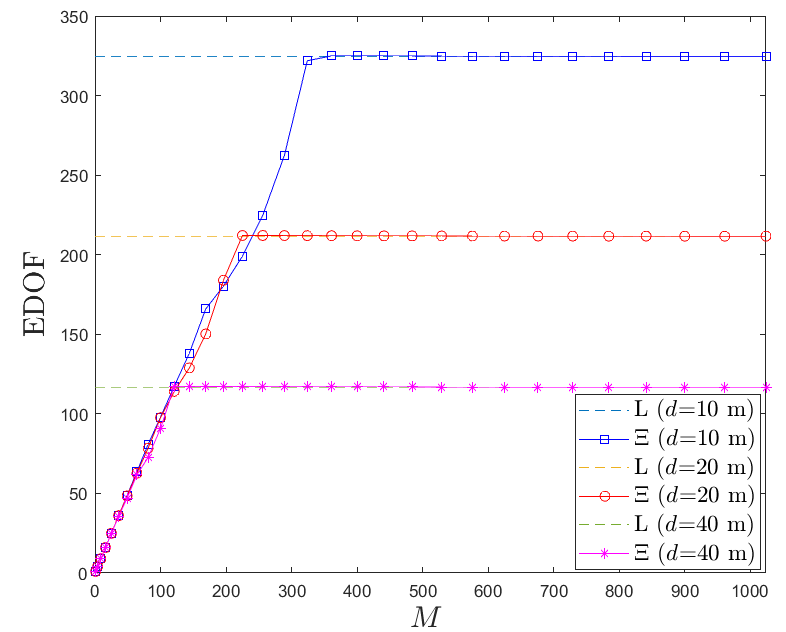}}
  \caption{\textcolor{black}{Change of EDOF with the increase of the number of sampling points $M$ in the domain $D$.}}
  \label{edof}
\end{figure}

\begin{figure}[t]
  \centering
\centerline{\includegraphics[width=8.4cm,height=6.5cm]{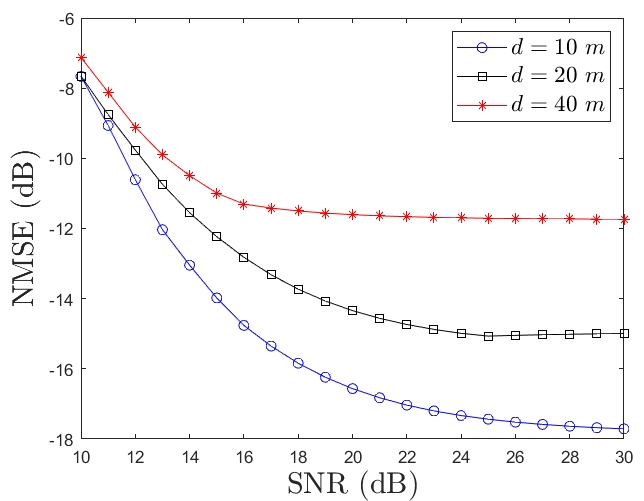}}
  \caption{\textcolor{black}{NMSE of PRCD versus SNR at the receiver with $K=32$.}}
  \label{nmse1}
\end{figure}

\begin{figure*}[t]
  \centering
\begin{minipage}[t]{0.32\linewidth}
\subfigure[Target relative permittivity]{
\includegraphics[width=6cm,height=5.5cm]{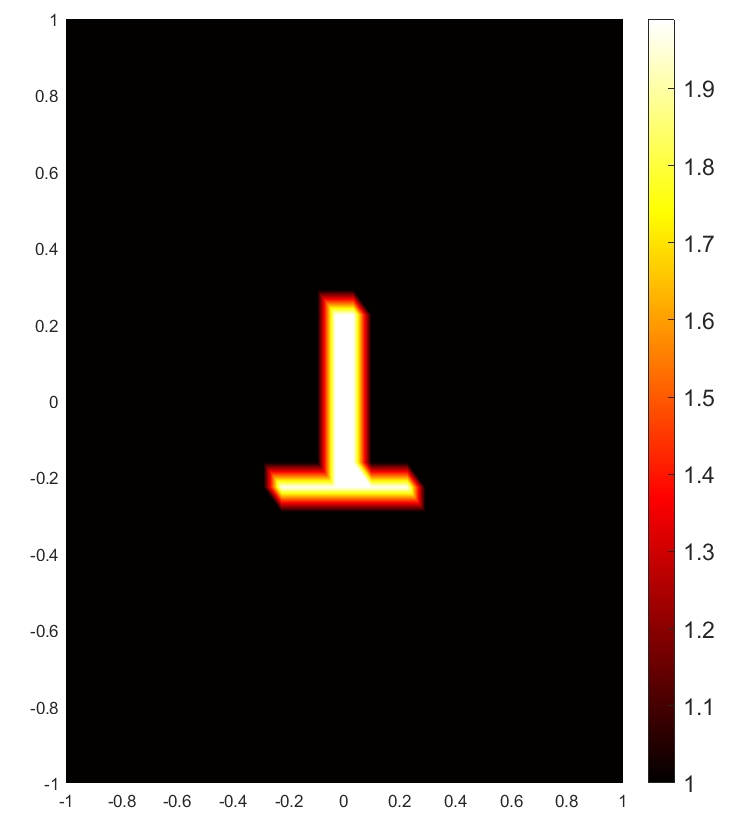}} 
\end{minipage}
\begin{minipage}[t]{0.32\linewidth}
\subfigure[Reconstrcted relative permittivity \protect\\ with SNR = 10 dB]{
\includegraphics[width=6cm,height=5.5cm]{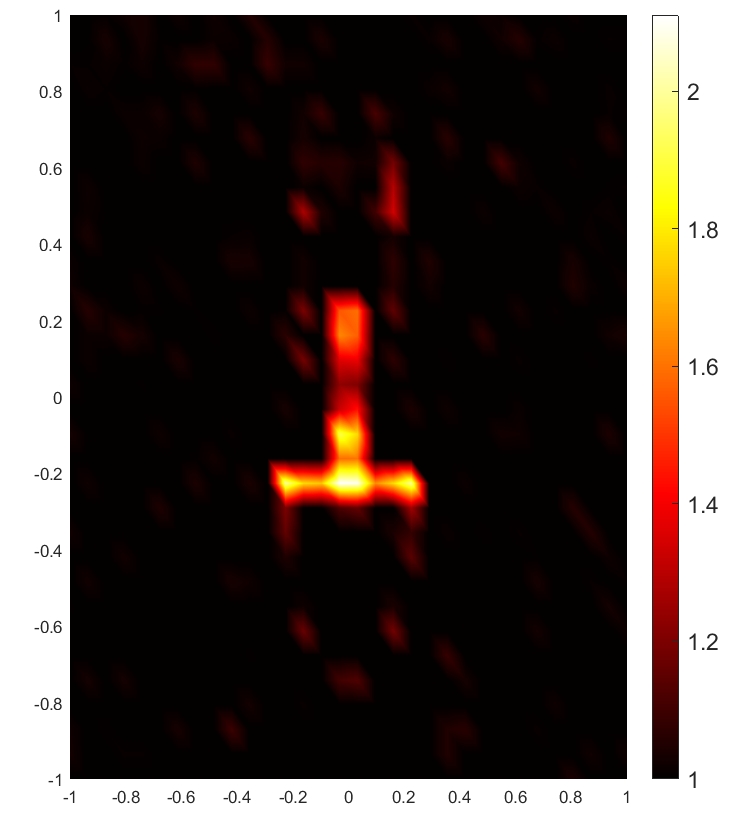}}  
\end{minipage}  
\begin{minipage}[t]{0.32\linewidth}
\subfigure[Reconstrcted relative permittivity \protect\\ with SNR = 30 dB]{
\includegraphics[width=6cm,height=5.5cm]{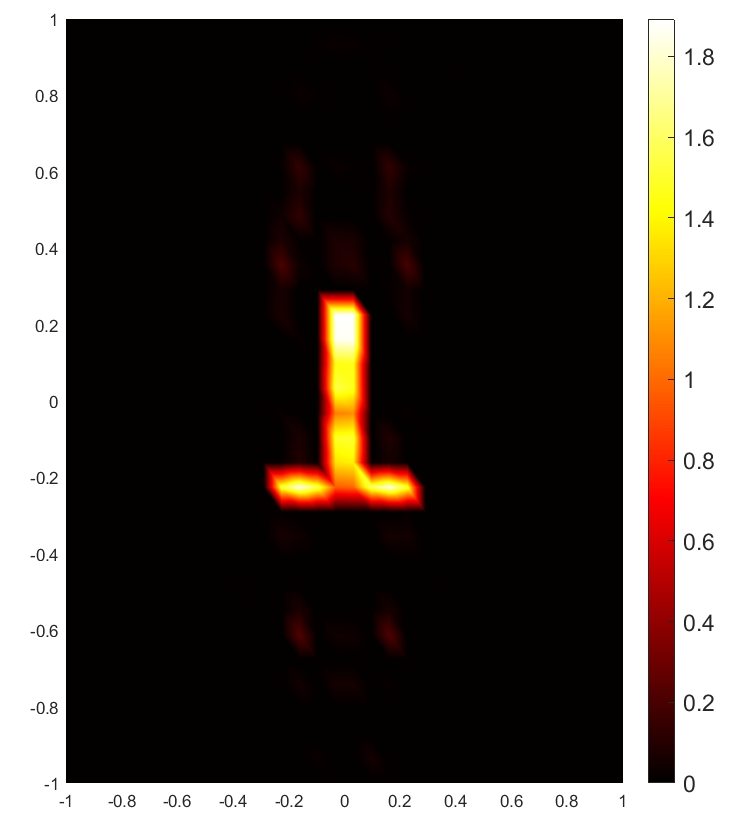}}   
\end{minipage} \\              
\begin{minipage}[t]{0.32\linewidth}
\subfigure[Target conductivity]{
\includegraphics[width=6cm,height=5.5cm]{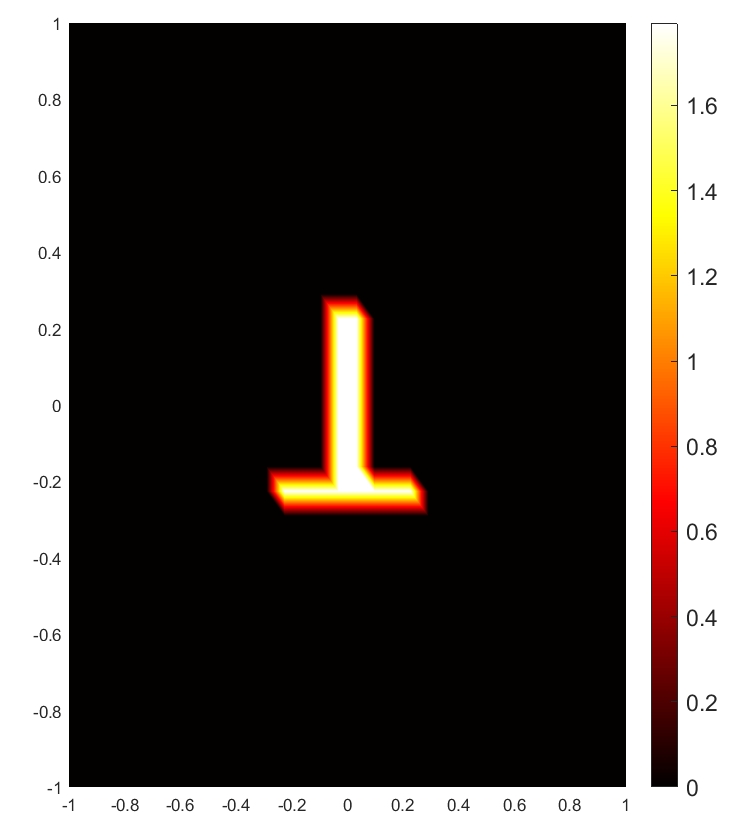}}
\end{minipage}
\begin{minipage}[t]{0.32\linewidth}
\subfigure[Reconstrcted conductivity   with SNR = 10 dB]{
\includegraphics[width=6cm,height=5.5cm]{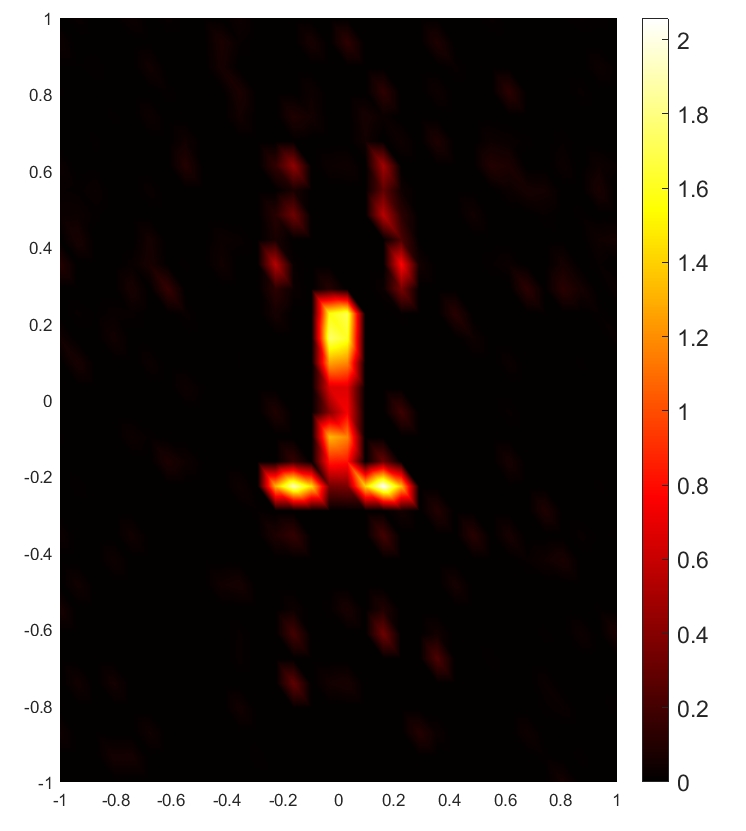}}
\end{minipage}
\begin{minipage}[t]{0.32\linewidth}
\subfigure[Reconstrcted conductivity  with SNR = 30 dB]{
\includegraphics[width=6cm,height=5.5cm]{30dB_conductivity.png}}
\end{minipage}
\caption{RPCD reconstruction results versus SNR with $d=20$ m and $K=32$. 
Unit of conductivity is mS/m.
}
\label{image1}
\end{figure*}

Correspondingly, the change of mutual coherence of the general sensing matrix $\tilde{\mathbf{E}}$ with the increase of the number of subcarriers $K$ is demonstrated in Fig.~\ref{mut}. 
It is seen that $\mu$ decreases with the increase of $K$, which indicates that the multi-frequency scheme can enhance the EM property sensing performance. 
When $K$ reaches a certain threshold, $\mu$ scarcely changes, where the transmitter-target 
distance $d$ is the decisive factor of sensing performance.
When the distance becomes larger, the mutual coherence of $\tilde{\mathbf{E}}$ also becomes larger, which indicates worse sensing quality.

In order to demonstrate the change of EDOF for the transmitter-target channel with the increase of the number of sampling points in the domain $D$, we show the EDOF of the central-frequency subcarrier with the increase of $M$ in Fig.~\ref{edof}.  
We define $L$ as the limit value of $\Xi$ when $M$ approaches infinity in (\ref{dis}).
EDOF increases with the increase of $M$ almost linearly until it reaches an upper bound.
As shown in Fig.~\ref{edof}, when the distance between the target and the transmitter becomes larger, the upper bound of EDOF is smaller and is reached at a smaller $M$, 
indicating that the RPCD reconstruction quality is worse.



\subsection{RPCD Reconstruction Performance versus SNR}


We investigate NMSE of reconstructed RPCD versus the SNR at the receiver, as shown in Fig.~\ref{nmse1}. 
We set the number of subcarriers to be $K=32$ and
the distance between the target and the transmitter to be $d = 10, 20$, and $40$~m, respectively. 
It is seen from Fig.~\ref{nmse1} that NMSE decreases with the increase of SNR for all $d$, and the best performance is achieved when $d = 10$~m.
When SNR is smaller than $12$~dB, the NMSE is large and decreases slowly with the increase of SNR.  
In this stage, the differences among the NMSE for different $d$ are not pronounced. 
The NMSE decreases rapidly when SNR increases from $15$~dB to $20$~dB.  
When SNR reaches certain thresholds, the NMSE decreases to the error floor and hardly changes. 
At this point, the restriction of RPCD reconstruction quality is no longer the noise at the receiver, but is the imperfect condition of the mutual coherence of the sensing matrix.  
As the target gets farther away from the transmitter,
the mutual coherence of the sensing matrix becomes larger, 
leading to a larger error floor in RPCD reconstruction.
Moreover, an error floor is reached at lower SNR levels. 

To illustrate the RPCD reconstruction results vividly, we present the reconstructed images of the target 
based on relative permittivity (or conductivity) when $d = 20$~m and $K=32$ for SNR $=10$ dB and $30$ dB, respectively.
As shown in Fig.~\ref{image1}, both relative permittivity and conductivity reconstructed distribution can reproduce the general shape of the target. 
The RPCD reconstructed at SNR $=30$ dB is more accurate to demonstrate the target's shape compared to the RPCD reconstructed at SNR $=10$ dB.
Moreover, a higher SNR value results in more accurate reconstructed values of relative permittivity and conductivity, which leads to better representation of the target's real EM property.  

\begin{figure}[t]
  \centering
\centerline{\includegraphics[width=8.4cm,height=6.5cm]{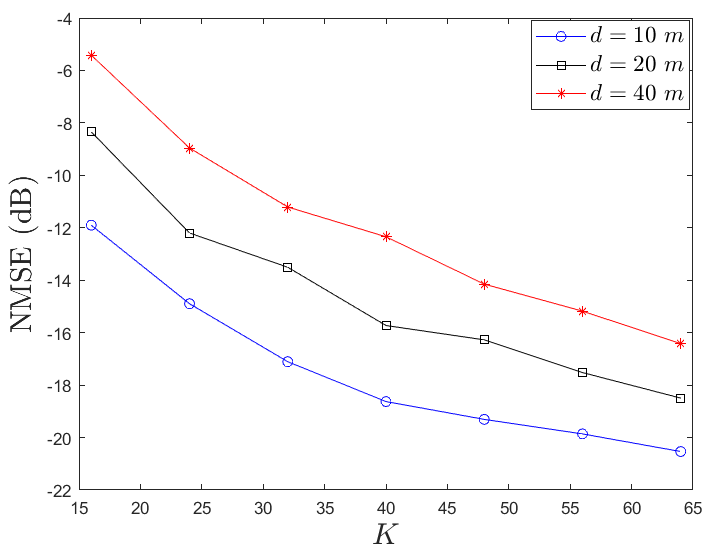}}
  \caption{\textcolor{black}{NMSE of PRCD versus $K$ with SNR = 30 dB.}}
  \label{nmse_K}
\end{figure}

\begin{figure*}[t]
\captionsetup[subfigure]{singlelinecheck=false}
\begin{minipage}[t]{0.32\linewidth}
\captionsetup[subfigure]{singlelinecheck=false}
\subfigure[Target relative permittivity]{
\includegraphics[width=6cm,height=5.5cm]{real_permittivity.png}} 
\end{minipage}
\begin{minipage}[t]{0.32\linewidth}
\subfigure[Reconstructed relative permittivity with $K$ = 16]{
\includegraphics[width=5.8cm,height=5.5cm]{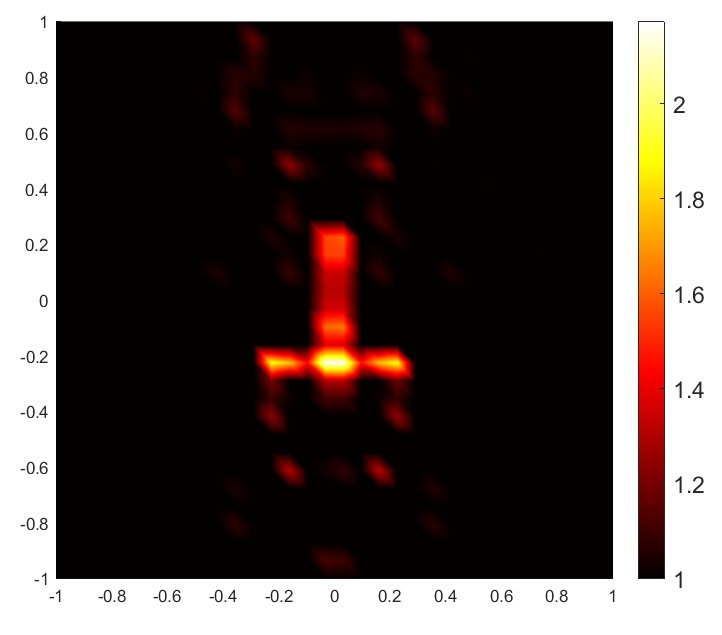}}  
\end{minipage}  
\begin{minipage}[t]{0.32\linewidth}
\subfigure[Reconstructed relative permittivity  with $K$ = 64]{
\includegraphics[width=6cm,height=5.5cm]{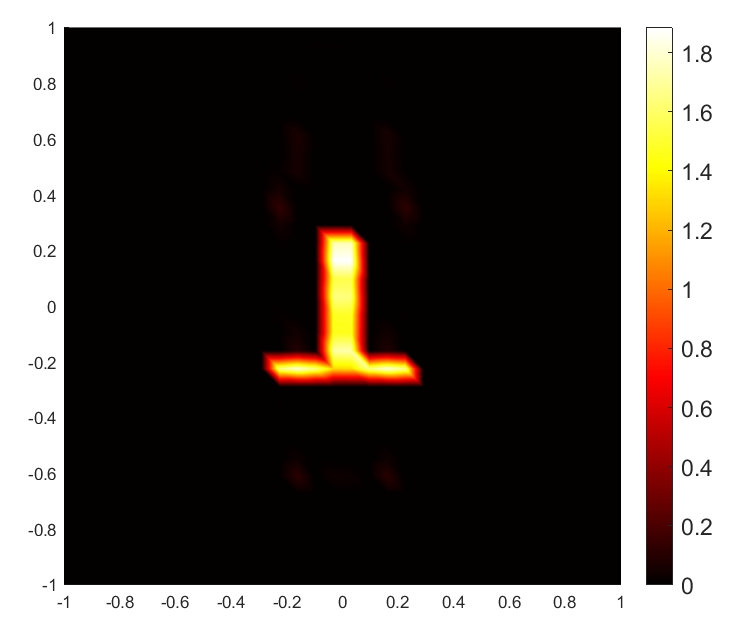}}   
\end{minipage} \\              
\begin{minipage}[t]{0.32\linewidth}
\captionsetup{singlelinecheck=false}
\subfigure[Target conductivity]{
\includegraphics[width=6cm,height=5.5cm]{real_conductivity.png}}
\end{minipage}
\begin{minipage}[t]{0.32\linewidth}
\subfigure[Reconstructed conductivity with $K$ = 16]
{\includegraphics[width=6cm,height=5.5cm]{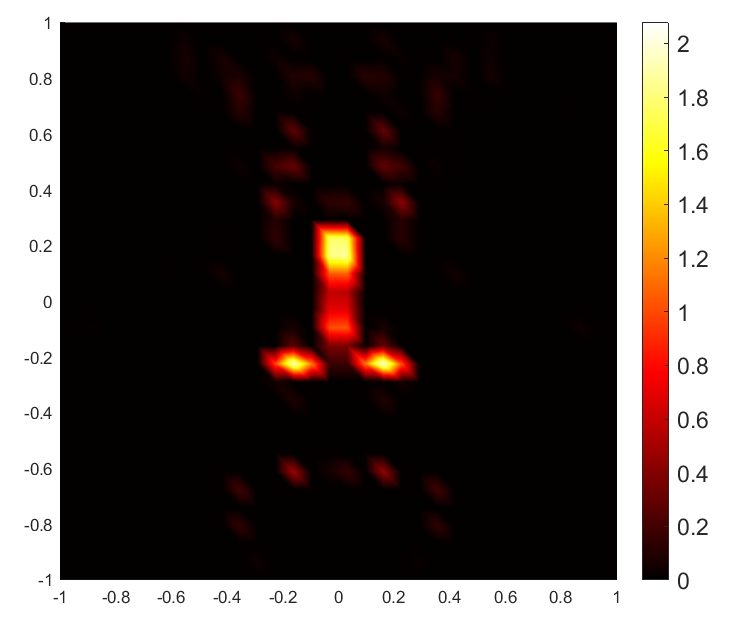}}
\end{minipage}
\begin{minipage}[t]{0.32\linewidth}
\subfigure[Reconstructed conductivity with $K$ = 64]{
\includegraphics[width=6cm,height=5.5cm]{64K_conductivity.png}}
\end{minipage}
\caption{RPCD reconstruction results versus $K$ with $d=20$ m and SNR = 30 dB. 
Unit of conductivity is mS/m.
}
\label{image_K}
\end{figure*}

\subsection{RPCD Reconstruction Performance versus Bandwidth}


We investigate NMSE of reconstructed RPCD versus the bandwidth by changing the number of subcarriers $K$ while keeping the frequency step unchanged, as shown in Fig.~\ref{nmse_K}. 
We set SNR = 30 dB and 
the distance between the target and the transmitter to be $d = 10$, 20, and $40$~m, respectively. 
It is seen from Fig.~\ref{nmse_K} that, NMSE decreases with the increase of $K$ for all $d$, and the best performance is achieved when $d = 10$~m. 
As the target gets farther from the transmitter, 
the mutual coherence of the sensing matrix becomes larger, causing larger errors in RPCD reconstruction.
Although the mutual coherence of the general sensing matrix $\tilde{\mathbf{E}}$ do not change significantly in $K\in[16,64]$ according to Fig.~\ref{mut}, the increase of $K$ enlarges the row number of $\tilde{\mathbf{E}}$ and leads to more measurement data, which improves the accuracy of RPCD reconstruction.



To illustrate the RPCD reconstruction results vividly, we present the reconstructed images of the target based on relative permittivity (or conductivity) with $d = 20$~m and SNR = 30 dB for $K=16$ and $64$, respectively.
As shown in Fig.~\ref{image_K}, the RPCD reconstructed with $K=64$ is more accurate to demonstrate the target's shape compared to the RPCD reconstructed at $K=16$.
Moreover, a larger number of subcarriers results in more accurate reconstructed values of relative permittivity and conductivity, which leads to the better representation of the target's real EM property.


\subsection{Material Classification Accuracy versus SNR}
\begin{figure}[t]
  \centering  \centerline{\includegraphics[width=8.4cm,height=6.5cm]{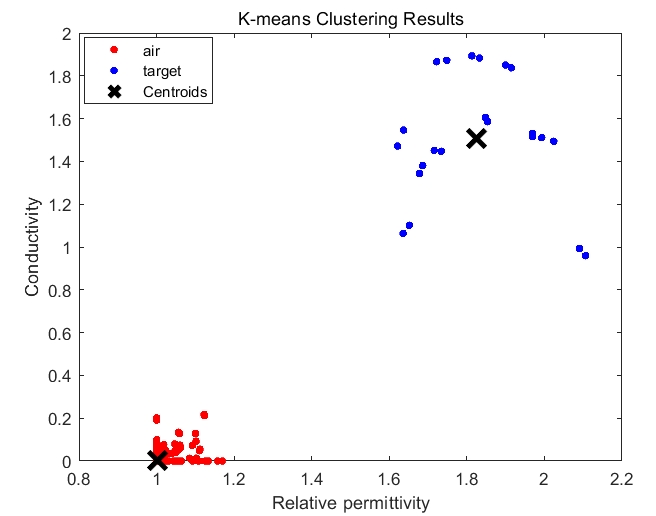}}  
  \caption{\textcolor{black}{Classification between the air and the target using the K-means clustering method.}}
  \label{kmeans}
\end{figure}

\begin{figure}[t]
  \centering  \centerline{\includegraphics[width=8.4cm,height=6.5cm]{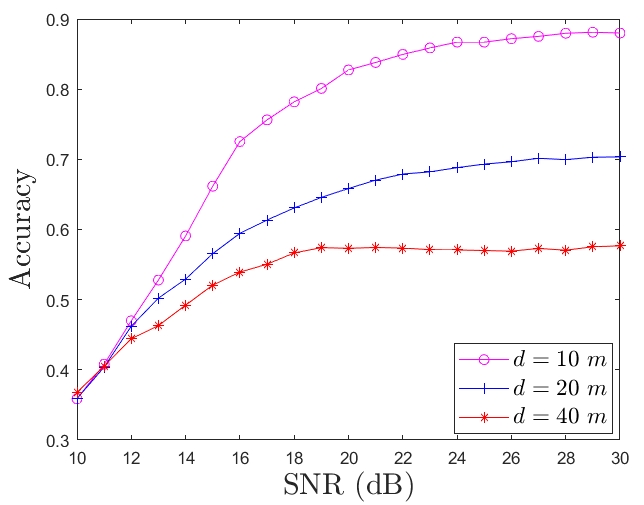}}  
  \caption{\textcolor{black}{Material classification accuracy versus SNR.}}
  \label{nmse2}
\end{figure}

Suppose the target may be composed of $10$ possible kinds of homogenous materials, such as wood, concrete, etc, whose relative permittivity and conductivity are precisely known. 
The shape of the target is set to be identical to the ``T" shape in Fig.~\ref{image1}, while the EM property of the target is decided by the material.

To provide an example of classification between the air and the target, 
we show the K-means clustering results when $K=32$, SNR = $20$ dB and $d=20$~m in Fig.~\ref{kmeans}.
In Fig.~\ref{kmeans}, each data point represents a sampling point in the domain $D$, and the colors of the data points represent the classification results. 
It is seen that, the cluster centroid of the air, representing the average permittivity and conductivity values of the air, is close to the $(1,0)$ point. 
On the other hand, the cluster centroid of the target material, representing its average permittivity and conductivity, is far away from the $(1,0)$ point.  
Thus, the sampling points occupied by air and those occupied by the target can be distinguished.

Using the sampling points occupied by the target, the material classification accuracy versus SNR with different $d$ when $K=32$ is shown in Fig.~\ref{nmse2}. 
We conduct 10000 Monte Carlo experiments for each material under each simulation setting. 
It is seen in Fig.~\ref{nmse2} that, the classification accuracy increases with the increase of SNR for all $d$, and the best performance is achieved with $d=10$~m.
When SNR is smaller than $12$~dB, the accuracy is smaller than $50\%$, and the differences among the classification accuracies with different $d$ are not pronounced. 
The accuracy increases rapidly when SNR increases from $10$~dB to $20$~dB and reaches an upper bound at a certain threshold of SNR.  

\subsection{Material Classification Accuracy versus Bandwidth}

\begin{figure}[t]
  \centering  \centerline{\includegraphics[width=8.4cm,height=6.5cm]{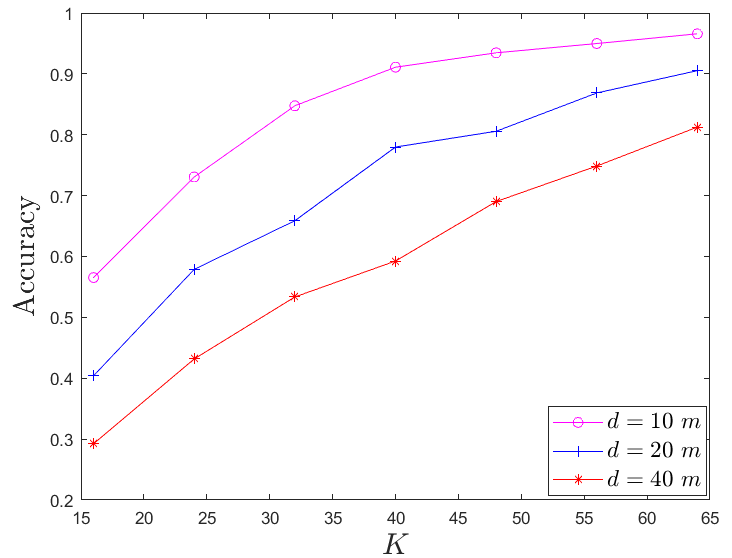}}  
  \caption{\textcolor{black}{Material classification accuracy versus $K$.}}
  \label{nmse2_K}
\end{figure}

Under the same simulation setting as in the previous subsection, we investigate the material classification accuracy versus the bandwidth by changing the number of subcarriers $K$ while keeping the frequency step unchanged, as shown in Fig.~\ref{nmse2_K}. 
We set SNR = 30 dB and $d=10, 20,$ and $40$~m respectively.
It is seen that, the classification accuracy increases with the increase of $K$ for all $d$. 
The best performance is achieved with $d=10$~m and $K=64$, where the accuracy is $96.7\%$.  
When $K$ is larger than $32$, the accuracies with all $d$ are larger than $50\%$. 
Generally speaking, the larger $d$ leads to 
the lower RPCD reconstruction error, which further results in the higher classification accuracy according to Fig.~\ref{nmse_K} and Fig.~\ref{nmse2_K}.

\section{Conclusion}
In this paper, we propose a groundbreaking EM property sensing and material identification scheme in ISAC systems by utilizing OFDM pilot signals. 
We develop an end-to-end EM propagation model grounded in Maxwell equations, enabling the reconstruction of relative permittivity and conductivity distributions within a defined sensing region. 
We then propose a multi-frequency EM property sensing method, employing compressive sensing techniques, and involving the optimization of the beamforming matrices.
Simulation results demonstrate that the proposed method is capable of achieving high-quality RPCD reconstruction and accurate material classification. 
This paper opens new possibilities for precise EM property sensing in ISAC systems, with implications for applications demanding advanced material identification capabilities.

 \small 
 \bibliographystyle{ieeetr}
 \bibliography{IEEEabrv,mainbib}

\begin{thebibliography}{10}

\bibitem{nature1}
W.~Gro{\ss}mann, H.~Horn, and O.~Niggemann, ``Improving remote material classification ability with thermal imagery,'' {\em Scientific Reports}, vol.~12, no.~1, p.~17288, 2022.

\bibitem{infrared1}
J.~L. Miller, {\em Principles of infrared technology}.
\newblock Springer, 1994.

\bibitem{isac8}
P.~Gao, L.~Lian, and J.~Yu, ``Cooperative {ISAC} with direct localization and rate-splitting multiple access communication: A pareto optimization framework,'' {\em IEEE J. Sel. Areas Commun.}, vol.~41, no.~5, pp.~1496--1515, 2023.

\bibitem{isac9}
Q.~Zhang, H.~Sun, X.~Gao, X.~Wang, and Z.~Feng, ``Time-division {ISAC} enabled connected automated vehicles cooperation algorithm design and performance evaluation,'' {\em IEEE J. Sel. Areas Commun.}, vol.~40, no.~7, pp.~2206--2218, 2022.

\bibitem{isac7}
X.~Cheng, D.~Duan, S.~Gao, and L.~Yang, ``Integrated sensing and communications ({ISAC}) for vehicular communication networks ({VCN}),'' {\em IEEE Internet Things J.}, vol.~9, no.~23, pp.~23441--23451, 2022.

\bibitem{gaoisac}
F.~Gao, L.~Xu, and S.~Ma, ``Integrated sensing and communications with joint beam-squint and beam-split for mmwave/thz massive mimo,'' {\em IEEE Trans. Commun.}, 2023.

\bibitem{isac1}
Z.~Ren, Y.~Peng, X.~Song, Y.~Fang, L.~Qiu, L.~Liu, D.~W.~K. Ng, and J.~Xu, ``Fundamental {CRB}-rate tradeoff in multi-antenna {ISAC} systems with information multicasting and multi-target sensing,'' {\em IEEE Trans. Wireless Commun.}, 2023.

\bibitem{mypaper3}
Y.~Jiang, F.~Gao, Y.~Liu, S.~Jin, and T.~Cui, ``Near field computational imaging with {RIS} generated virtual masks,'' 2023.

\bibitem{isac2}
F.~Dong, F.~Liu, Y.~Cui, W.~Wang, K.~Han, and Z.~Wang, ``Sensing as a service in 6{G} perceptive networks: A unified framework for {ISAC} resource allocation,'' {\em IEEE Trans. Wireless Commun.}, 2022.

\bibitem{isac3}
X.~Meng, F.~Liu, C.~Masouros, W.~Yuan, Q.~Zhang, and Z.~Feng, ``Vehicular connectivity on complex trajectories: Roadway-geometry aware isac beam-tracking,'' {\em IEEE Trans. Wireless Commun.}, 2023.

\bibitem{isac4}
N.~Su, F.~Liu, and C.~Masouros, ``Sensing-assisted eavesdropper estimation: An {ISAC} breakthrough in physical layer security,'' {\em IEEE Trans. Wireless Commun.}, 2023.

\bibitem{isac5}
K.~Chen, C.~Qi, O.~A. Dobre, and G.~Y. Li, ``Simultaneous beam training and target sensing in {ISAC} systems with {RIS},'' {\em IEEE Trans. Wireless Commun.}, 2023.

\bibitem{isac6}
Z.~He, W.~Xu, H.~Shen, D.~W.~K. Ng, Y.~C. Eldar, and X.~You, ``Full-duplex communication for {ISAC}: Joint beamforming and power optimization,'' {\em IEEE J. Sel. Areas Commun.}, 2023.

\bibitem{isac10}
S.~Li, W.~Yuan, C.~Liu, Z.~Wei, J.~Yuan, B.~Bai, and D.~W.~K. Ng, ``A novel {ISAC} transmission framework based on spatially-spread orthogonal time frequency space modulation,'' {\em IEEE J. Sel. Areas Commun.}, vol.~40, no.~6, pp.~1854--1872, 2022.

\bibitem{RIS_image}
Y.~Tao and Z.~Zhang, ``Distributed computational imaging with reconfigurable intelligent surface,'' in {\em 2020 International Conference on Wireless Communications and Signal Processing (WCSP)}, pp.~448--454, Oct. 2020.

\bibitem{mypaper}
Y.~Jiang, F.~Gao, M.~Jian, S.~Zhang, and W.~Zhang, ``Reconfigurable intelligent surface for near field communications: Beamforming and sensing,'' {\em IEEE Trans. Wireless Commun.}, vol.~22, no.~5, pp.~3447--3459, 2023.

\bibitem{twin}
A.~Alkhateeb, S.~Jiang, and G.~Charan, ``Real-time digital twins: Vision and research directions for 6g and beyond,'' {\em IEEE Communications Magazine}, 2023.

\bibitem{twin2}
Y.~Cui, W.~Yuan, Z.~Zhang, J.~Mu, and X.~Li, ``On the physical layer of digital twin: An integrated sensing and communications perspective,'' {\em IEEE J. Sel. Areas Commun.}, 2023.

\bibitem{lipp}
J.~O. Vargas and R.~Adriano, ``Subspace-based conjugate-gradient method for solving inverse scattering problems,'' {\em IEEE Trans. Antennas Propag.}, vol.~70, no.~12, pp.~12139--12146, 2022.

\bibitem{m-born}
T.-F. Wei, X.-H. Wang, L.~Wang, Z.~Feng, and B.-Z. Wang, ``Efficient born iterative method for inverse scattering based on modified forward-solver,'' {\em IEEE Access}, vol.~8, pp.~229101--229107, 2020.

\bibitem{v-born}
N.~Zaiping, Y.~Feng, Z.~Yanwen, and Z.~Yerong, ``Variational born iteration method and its applications to hybrid inversion,'' {\em IEEE Trans. Geosci. Remote Sens.}, vol.~38, no.~4, pp.~1709--1715, 2000.

\bibitem{operator}
Z.~Liu and Z.~Nie, ``Subspace-based variational born iterative method for solving inverse scattering problems,'' {\em IEEE Geoscience and Remote Sensing Letters}, vol.~16, no.~7, pp.~1017--1020, 2019.

\bibitem{operator0}
J.~O. Vargas and R.~Adriano, ``Subspace-based conjugate-gradient method for solving inverse scattering problems,'' {\em IEEE Transactions on Antennas and Propagation}, vol.~70, no.~12, pp.~12139--12146, 2022.

\bibitem{MOM}
Y.~Ren, Q.~H. Liu, and Y.~P. Chen, ``A hybrid {FEM}/{MoM} method for 3-{D} electromagnetic scattering in layered medium,'' {\em IEEE Trans. Antennas Propag.}, vol.~64, no.~8, pp.~3487--3495, 2016.

\bibitem{FEM}
F.~M. Kahnert, ``Numerical methods in electromagnetic scattering theory,'' {\em Journal of Quantitative Spectroscopy and Radiative Transfer}, vol.~79, pp.~775--824, 2003.

\bibitem{FDTD}
W.~Hou, M.~Azadifar, M.~Rubinstein, Q.~Zhang, and F.~Rachidi, ``An efficient fdtd method to calculate lightning electromagnetic fields over irregular terrain adopting the moving computational domain technique,'' {\em IEEE Trans. Electromagn. Compat.}, vol.~62, no.~3, pp.~976--980, 2019.

\bibitem{lipp2}
F.-F. Wang and Q.~H. Liu, ``A hybrid {B}orn iterative {B}ayesian inversion method for electromagnetic imaging of moderate-contrast scatterers with piecewise homogeneities,'' {\em IEEE Trans. Antennas Propag.}, vol.~70, no.~10, pp.~9652--9661, 2022.

\bibitem{cs2}
C.~Stoeckle, J.~Munir, A.~Mezghani, and J.~A. Nossek, ``Do{A} estimation performance and computational complexity of subspace-and compressed sensing-based methods,'' in {\em WSA 2015; 19th International ITG Workshop on Smart Antennas}, pp.~1--6, VDE, 2015.

\bibitem{born}
D.~Tajik, R.~Kazemivala, and N.~K. Nikolova, ``Real-time imaging with simultaneous use of {B}orn and {R}ytov approximations in quantitative microwave holography,'' {\em IEEE Trans. Microw. Theory Tech.}, vol.~70, no.~3, pp.~1896--1909, 2021.

\bibitem{born2}
J.~Guillemoteau, P.~Sailhac, and M.~Behaegel, ``Fast approximate {2D} inversion of airborne tem data: {B}orn approximation and empirical approach,'' {\em Geophysics}, vol.~77, no.~4, pp.~WB89--WB97, 2012.

\bibitem{kmeans}
M.~Ahmed, R.~Seraj, and S.~M.~S. Islam, ``The {K}-means algorithm: A comprehensive survey and performance evaluation,'' {\em Electronics}, vol.~9, no.~8, p.~1295, 2020.

\bibitem{mahalanobis}
E.~Bayram and V.~Nabiyev, ``Image segmentation by using k-means clustering algorithm in euclidean and mahalanobis distance calculation in camouflage images,'' in {\em 2020 28th Signal Processing and Communications Applications Conference (SIU)}, pp.~1--4, IEEE, 2020.

\bibitem{kmeans_complexity}
X.~Jin and J.~Han, {\em K-Means Clustering}, pp.~563--564.
\newblock Boston, MA: Springer US, 2010.

\bibitem{RIP}
E.~Candes and T.~Tao, ``Decoding by linear programming,'' {\em IEEE Trans. Inf. Theory}, vol.~51, no.~12, pp.~4203--4215, 2005.

\bibitem{adaptive}
B.~Kilic, A.~Güngör, M.~Kalfa, and O.~Arıkan, ``Adaptive measurement matrix design in direction of arrival estimation,'' {\em IEEE Trans. Signal Process.}, vol.~70, pp.~4742--4756, 2022.

\bibitem{compressed_sensing}
D.~Donoho, ``Compressed sensing,'' {\em IEEE Trans. Inf. Theory}, vol.~52, no.~4, pp.~1289--1306, 2006.

\bibitem{adaptive2}
T.~Huang, Y.~Liu, H.~Meng, and X.~Wang, ``Adaptive compressed sensing via minimizing cramer–rao bound,'' {\em IEEE Signal Process. Lett.}, vol.~21, no.~3, pp.~270--274, 2014.

\bibitem{SDR}
Z.-q. Luo, W.-k. Ma, A.~M.-c. So, Y.~Ye, and S.~Zhang, ``Semidefinite relaxation of quadratic optimization problems,'' {\em IEEE Signal Processing Magazine}, vol.~27, no.~3, pp.~20--34, 2010.

\bibitem{low_rank_theorem}
G.~H. Golub, A.~Hoffman, and G.~W. Stewart, ``A generalization of the {E}ckart-{Y}oung-{M}irsky matrix approximation theorem,'' {\em Linear Algebra and its applications}, vol.~88, pp.~317--327, 1987.

\bibitem{low_rank}
A.~Dax {\em et~al.}, ``Low-rank positive approximants of symmetric matrices,'' {\em Advances in Linear Algebra \& Matrix Theory}, vol.~4, no.~03, p.~172, 2014.

\bibitem{myEDOF}
Y.~Jiang and F.~Gao, ``Electromagnetic channel model for near field {MIMO} systems in the half space,'' {\em IEEE Commun. Lett.}, vol.~27, no.~2, pp.~706--710, 2023.

\bibitem{1}
T.~A. Sleasman, M.~F. Imani, A.~V. Diebold, M.~Boyarsky, K.~P. Trofatter, and D.~R. Smith, ``Implementation and characterization of a two-dimensional printed circuit dynamic metasurface aperture for computational microwave imaging,'' {\em IEEE Trans. Antennas Propagat.}, vol.~69, no.~4, pp.~2151--2164, Apr. 2021.

\bibitem{2}
G.~Sun, R.~He, B.~Ai, Z.~Ma, P.~Li, Y.~Niu, J.~Ding, D.~Fei, and Z.~Zhong, ``A 3{D} wideband channel model for {RIS}-assisted {MIMO} communications,'' {\em IEEE Trans. Vehi. Tech.}, vol.~71, no.~8, pp.~8016--8029, Aug. 2022.

\bibitem{9}
O.~Rinchi, A.~Elzanaty, and M.-S. Alouini, ``Compressive near-field localization for multipath {RIS}-aided environments,'' {\em IEEE Commun. Lett.}, vol.~26, no.~6, pp.~1268--1272, Aug. 2022.

\end{thebibliography}

\ifCLASSOPTIONcaptionsoff
  \newpage
\fi



%

\end{document}